\documentclass[preprint]{revtex4-1}

\usepackage{graphicx}% Include figure files
\usepackage{bm}% bold math
\usepackage{scalerel}
\usepackage[utf8]{inputenc}
\usepackage[T1]{fontenc}
\usepackage{mathptmx}
\usepackage{xcolor}
\usepackage{amsmath}
\usepackage[colorlinks=true, citecolor=blue,
  linkcolor=blue]{hyperref}

\begin{document}

%\preprint{AIP/123-QED}

\title{Dynamics of an internally actuated weakly elastic sphere in a general quadratic flow\\}
% Force line breaks with \\

\author{Shashikant Verma}
\affiliation{%
Department of Mechanical Engineering, Indian Institute of Technology, Ropar 140001, India
%NORDITA, Roslagstullsbacken 23, SE-10691 Stokcholm, Sweden%\\This line break forced% with \\
}%

\author{Navaneeth K. Marath}
% \homepage{http://www.Second.institution.edu/~Charlie.Author.}
\affiliation{%
Department of Mechanical Engineering, Indian Institute of Technology, Ropar 140001, India
}%

%\affiliation{%
%Mathematical Institute, University of Oxford, Oxford OX2 6GG, United Kingdom
%}%
\date{\today}% It is always \today, today,

\begin{abstract}
Internally actuated elastic particles are widely used in biomedical applications. It is imperative to understand the dynamics of such particles in pressure-driven microfluidic devices to manipulate their motion. We analytically examine the dynamics of an internally actuated elastic particle translating in a general unbounded quadratic flow in the inertialess limit. We consider the particle as a compressible weakly elastic sphere, and its motion is controlled by applying an external point force and a point torque at the centre of its undeformed shape. The fluid and the particle are modelled using the Stokes and the Navier elasticity equations, respectively. We use the domain perturbation method to capture the particle deformation. The point force and the point torque are obtained until \textit{O}($\alpha^2$), assuming $\alpha\ll 1$. Here, $\alpha$ is the measure of the particle elastic strain induced due to the fluid viscous stress. We present the results for the particle motion in a general unbounded quadratic flow. The results are simplified further for the motion along the centreline in the quadratic component of three Poiseuille flows: 1) elliptical Poiseuille, 2) plane Poiseuille, and 3) Hagen-Poiseuille flows. In the general quadratic flow, the point force at \textit{O}($\alpha$) is aligned with the particle velocity, while the force at \textit{O}($\alpha^2$) acts at an angle to the velocity. Furthermore, the torque is non-zero due to elastic effects at \textit{O}($\alpha$) and \textit{O}($\alpha^2$). For all the three Poiseuille flows, the point force until \textit{O}($\alpha^2$) is aligned with the particle velocity, while the torque comes as zero.
\end{abstract}

\maketitle

% \begin{quotation}
% The ``lead paragraph'' is encapsulated with the \LaTeX\ 
% \verb+quotation+ environment and is formatted as a single paragraph before the first section heading. 
% (The \verb+quotation+ environment reverts to its usual meaning after the first sectioning command.) 
% Note that numbered references are allowed in the lead paragraph.
% %
% The lead paragraph will only be found in an article being prepared for the journal \textit{Chaos}.
% \end{quotation}

\section{Introduction}
\begin{figure*}
\centerline{\includegraphics[width=0.8\textwidth]{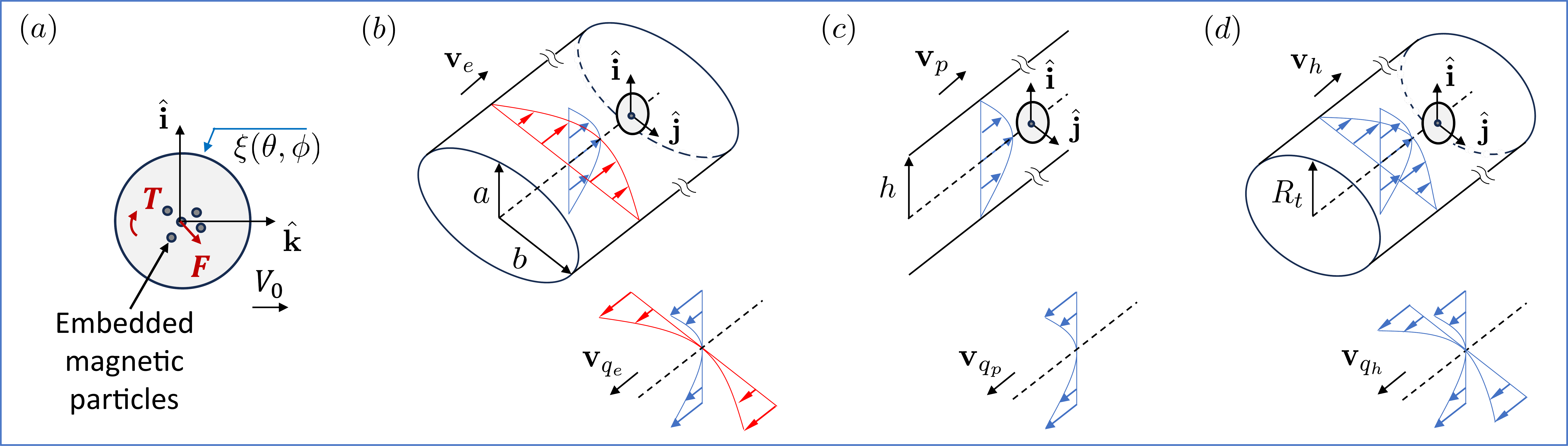}}
  \caption{(a) Schematic of $Fe_3O_4$ nanoparticles embedded in a polymer bead.\citep{peng2008magnetically} Schematic of an elastic sphere translating with velocity $V_0 \mathbf{\hat{k}}$ along the centerline of a channel: (b) inside a straight tube of elliptic cross-section, with $a$ and $b$ as the semi-minor and semi-major axes, respectively; elliptic Poiseuille flow ($\mathbf{v}_e$, top subfigure) and its quadratic component ($\mathbf{v}_{q_e}$, bottom subfigure), (c) between two infinite parallel plates separated by a distance $2h$; plane Poiseuille flow ($\mathbf{v}_p$, top subfigure) and its quadratic component ($\mathbf{v}_{q_p}$, bottom subfigure), and (d) inside a cylindrical tube of radius $R_t$; Hagen-Poiseuille flow ($\mathbf{v}_h$, top subfigure) and its quadratic component ($\mathbf{v}_{q_h}$, bottom subfigure). The sphere is subjected to an external point force ($\mathbf{F}$) and a point torque ($\mathbf{T}$).}
\label{fig:schematic_pp_hp_flow}
\end{figure*}

Internally actuated elastic particles are used in many biomedical applications. A notable example of such particles is magnetoresponsive polymer beads\citep{philippova2011magnetic} consisting of a polymer matrix embedded with magnetic nano- or microparticles (MPs). The polymer provides structural integrity and elasticity, while the embedded MPs impart magnetic responsiveness. These polymer beads are widely used in in vitro magnetic separation of biological cells\citep{philippova2011magnetic,seyfoori2023microfluidic} and in targeted drug delivery.\citep{sung2021magnetic} By binding to non-magnetic particles, magnetically responsive beads enable their controlled manipulation under externally applied magnetic fields.\citep{safarik1995application,safarikova2001application,philippova2011magnetic} The MPs are distributed either uniformly or non-uniformly within the polymer matrix.\citep{philippova2011magnetic} Fig. \ref{fig:schematic_pp_hp_flow}(a) presents a schematic of MPs embedded within a polymer bead. Each embedded magnetic particle responds to the external magnetic field as a localised force that internally actuates the particle. Additionally, MPs can also experience localised torques when subjected to a rotating magnetic field.\citep{moerland2019rotating} The localised forces and torques can be modelled as point forces and point torques, respectively. Such a model has been employed to analytically investigate the dynamics of a weakly elastic sphere (embedded with a single magnetic particle at its centre) translating parallel to a rigid wall.\citep{verma2025dynamics} The dynamics and deformed shape of a particle moving in a microchannel are determined by the force distribution acting on the particle. For instance, a body force distributed throughout the particle volume yields different dynamics than a localised body force. To analyse the dynamics of internally actuated particles in pressure-driven microfluidic devices, it is imperative to understand their behaviour in quadratic flows. Pressure-driven flows such as Poiseuille flows in elliptical channels, cylindrical tubes and between parallel plates are composed of uniform, linear and quadratic components. In particular, along the centreline of channels/tubes, the linear component vanishes and the particle that is translating with the local ambient velocity experiences only a quadratic component of the flow. Understanding the particle motion along the centreline is important in flow cytometry.\citep{cavett2023hydrogel,tobias2024development}

Several studies have analysed the dynamics of rigid\citep{wang2020dynamics,tai2020cross} and deformable particles\citep{yang1989motions,nadim1991motion} in quadratic flows. Theoretical analysis has been carried out to understand the motion and deformation of a spherical drop in quadratic flows, such as paraboloidal and stagnation flows,\citep{yang1989motions} and in nonlinear cubic shear flows.\citep{favelukis2022drop} Yang and Lee\citep{yang1989motions} have derived the generalised form of Faxen's law for a spherical drop immersed in an arbitrary Stokes flow. The nonlinearity of the cubic shear flow deforms the drop [viscosity ratio of \textit{O}(1)] and sometimes aligns it into two different orientations, at $\pm45^{\circ}$ to the flow. A deformable spherical drop in an unbounded Poiseuille flow exhibits lateral migration toward the channel centreline, depending on the product of the Hadamard–Rybczynski terminal settling velocity and the maximum Poiseuille flow velocity.\citep{haber1971dynamics} Stokes flow around a nearly spherical drop in a general quadratic flow has been determined using the vectorial form of Lamb’s general solution to the Stokes equations.\citep{nadim1991motion} Recently, the hydrodynamics and morphology of a spherical active compound drop have been analysed in the general quadratic flow.\citep{chaithanya2023active} The results from these studies\citep{nadim1991motion,chaithanya2023active} have shown that an initially spherical simple or compound drop deforms into a three-lobe shape under the influence of the hexapolar component of the quadratic flow.  Experimentally, it has been observed that the initial spherical simple drop deforms to approximately the three-lobe shape when it is subjected to the hexapolar quadratic flow.\citep{razzaghi2023deformation}   

While the dynamics of drops in the quadratic flow has been extensively studied, that of elastic particles remains less explored. The key difference between the behaviour of a drop and an elastic particle comes from the distinct constitutive and governing equations that describe their motion and deformation. Elastic particles as cell models are studied for their potential in diagnosing diseases through mechanical-property-dependent flow behaviour.\citep{hou2009deformability,villone2019dynamics} Further, predicting the internal stresses is important in cells such as white blood cells and osteoblasts, as the stresses can alter the expression of certain genes and, consequently, alter cellular properties.\citep {zhang2012gene,jin2020shear,finney2024impact} Modelling these cells as elastic particles helps in such predictions. Particle elasticity also plays a crucial role in drug delivery processes, as stiffer hydrogel particles are often internalised by immune cells more effectively than softer ones.\citep{anselmo2017impact}
One of the earlier works in modelling a particle as an elastic solid was carried out by Tam,\citep{tam1973transverse} who studied the dynamics of a compressible elastic sphere in an unbounded simple shear flow. Murata has captured the dynamics of a compressible elastic sphere, sedimenting in an unbounded quiescent fluid\citep{murata1980deformation} and also studied the dynamics of an incompressible elastic sphere in an arbitrary flow field,\citep{murata1981deformation} in the Stokes limit. 
Finney et al.\citep{finney2024impact} have examined the influence of channel confinement on the deformation of an incompressible elastic sphere, subjected to an axial body force and translating along the centreline of a rigid cylindrical tube using analytical and numerical techniques. They find that the deformed shape of the sphere can be bullet-like, an anti-bullet or spherical, depending on the strength of the applied body force relative to the viscous force.  
  
In the present work, we analyse the dynamics of an internally actuated weakly elastic spherical particle translating in a general unbounded quadratic flow. We assume that the particle is a homogeneous, isotropic, compressible Hookean solid and contains a centrally embedded magnetic particle. The motion of the particle is controlled by a localised force/torque, modelled as a point force/torque, induced by an external magnetic field. We use the domain perturbation method and series solutions to the governing equations of the particle and the fluid. First, we present the results for the particle translating in general quadratic flow and then the results are simplified for particle translating along the centreline in the quadratic component of three fully developed flows: 1) flow through the straight rigid tube of the elliptic cross section with $a$ and $b$ being the semi-minor and semi-major axes, respectively, called elliptical Poiseuille flow shown in Fig. \ref{fig:schematic_pp_hp_flow}(b), 2) flow between two infinite parallel rigid plates separated by a distance $2 h$ apart, called plane Poiseuille flow shown in Fig. \ref{fig:schematic_pp_hp_flow}(c), and 3) flow inside a straight rigid cylindrical tube of radius $R_t$, called Hagen-Poiseuille flow shown in Fig. \ref{fig:schematic_pp_hp_flow}(d).

We observe that, in general quadratic flow, the point force arising from elastic effects appears at \textit{O}($\alpha$) and is aligned with the particle velocity, whereas the force at \textit{O}($\alpha^2$) acts at an angle to it. However, in all three Poiseuille flows we considered, the force until \textit{O}($\alpha^2$) remains aligned with the particle velocity. The point torque is nonzero at both \textit{O}($\alpha$) and \textit{O}($\alpha^2$) in general quadratic flow but vanishes when the particle translates along the centreline in Poiseuille flows. The leading-order deformed shape of the particle is compared with that of a spherical active compound drop (active particle embedded inside the drop)\citep{chaithanya2023active} in plane Poiseuille flow. The compressible nature of the elastic particle leads to certain morphological differences between the elastic particle and the drop.

The paper is structured as follows. In Section \ref{sec:Mathematical formulation}, we present the governing equations for both the fluid and the particle, along with the quadratic component of a general ambient flow. We also discuss series solutions to the Stokes and Navier elasticity equations, as well as the domain perturbation method. In Section \ref{sec:results}, we present the results for the general quadratic flow and for the quadratic components of the elliptical Poiseuille, plane Poiseuille and Hagen-Poiseuille flows. We plot the streamlines near the particle and its deformed shape. The leading-order deformed shape of the particle in a plane Poiseuille flow is compared with a drop. Finally, we present the summary of the main results and key observations in Section \ref{sec:summary}. 
 
\section{Mathematical formulation}\label{sec:Mathematical formulation}
We consider a weakly elastic spherical particle of undeformed radius ($R_0$) moving in a general unbounded quadratic flow. An external point force and a point torque are applied at the centre of the undeformed particle to translate it at a velocity $V_0\mathbf{\hat{k}}$, as shown in Fig. \ref{fig:schematic_pp_hp_flow}(a). We assume that the viscous effects dominate over the inertial effects, i.e the Reynolds number, defined as $Re =\rho_fV_0 R_0/ \mu \ll 1$. Here, $\mu$ and $\rho_f$ are the fluid viscosity and density, respectively. The fluid is considered Newtonian and incompressible, and is governed by the Stokes equations given by
\begin{align}
 - \mathbf{\nabla} p+\mu \nabla^2  \mathbf{v} ={0},  \label{eq:stokes} \\
   \mathbf{\nabla}\cdot \mathbf{v}={0}. \label{eq:continuity}
 \end{align}
In eqn (\ref{eq:stokes}) and (\ref{eq:continuity}), $p$ and $\mathbf{v}$ are the pressure and velocity fields in the fluid, respectively. The particle is modelled using the Navier elasticity equations given by
\begin{align}
  \mathbf[\lambda+G]  \mathbf{\nabla}\left( \mathbf{\nabla} \cdot  \mathbf{u}\right)+G  \mathbf{\nabla}^2  \mathbf{u} +\mathbf{F}\delta(\mathbf{x})+\frac{1}{2}\left[\mathbf{\nabla}\delta(\mathbf{x})\times \mathbf{T} \right] ={0}.\label{eq:elasticity}
\end{align}
Here, $\lambda$ and $G$ are the Lam\'e's constants and $\mathbf{u}$ is the displacement field in the particle. The fluid stress ($\boldsymbol{\sigma}$) is given by
\begin{align}
 \boldsymbol{\sigma}=-p \mathbf{I}+\mu \left(\nabla \mathbf{v}+[\nabla \mathbf{v}]^T\right) \label{eq:fluid-stress constitutive}
\end{align}
and the solid stress ($\hat{\boldsymbol{\sigma}}$) is given by
\begin{align}
   \hat{\boldsymbol{\sigma}}=\lambda \left(\nabla\cdot\mathbf{u}\right)\mathbf{I}+G\left(\nabla \mathbf{u}+[\nabla \mathbf{u}]^T\right).\label{eq:solid-stress constitutive}
\end{align}
The variables in the present analysis are non-dimensionalised using the following scalings
\begin{align}
    \mathbf{v^*}=\frac{\mathbf{v}}{V_0};\,\,
    p^*=\frac{p R_0}{\mu V_0}; \,\,\boldsymbol{\sigma}^*=\frac{\boldsymbol{\sigma} R_0}{\mu V_0};\,\,
    \mathbf{u^*}=\frac{\mathbf{u}}{R_{0}};\,\,    \hat{\boldsymbol{\sigma}}^*=\frac{\hat{\boldsymbol{\sigma}}}{G}. \label{eq:nondimensional variables}
\end{align}
The ratio of the particle bulk modulus ($\lambda$) to its shear modulus ($G$) is represented by $\Gamma$; $\Gamma=\lambda/G$. Deformability of the particle can be expressed as a measure of elastic strain induced in the particle due to fluid viscous stress, given by $\alpha=\mu V_0/(G R_0)$.\citep{finney2024impact,verma2025dynamics} The parameter $\alpha$ can also be defined as an elastic capillary number, quantifying the ratio of viscous to elastic stress,\citep{villone2019dynamics} or as the ratio of viscous to elastic force.\citep{murata1980deformation, murata1981deformation, nasouri2017elastic} The non-dimensional variables in eqn (\ref{eq:nondimensional variables}) are denoted using an asterisk symbol, and the symbol is omitted hereafter for convenience.

To analyse the particle dynamics, we solve eqn (\ref{eq:stokes})-(\ref{eq:elasticity}) in a spherical coordinate system ($r$, $\theta$, $\phi$), whose origin coincides with the centre of the undeformed particle. The solutions are expressed in terms of the non-dimensional radial coordinate ( $\xi = r/R_0$), the polar ($\theta$) and the azimuthal ($\phi$) angles. The particle is assumed to deform to a steady shape with surface radius given by
\begin{align}
  \xi=1+f(\theta,\phi). \label{eq:surface}
\end{align}
Here, $f$ quantifies the surface deformation and depends on both the polar and azimuthal angles. The displacement of a material point on the deformed surface, relative to its position on the undeformed surface, is expressed in terms of $f$ as 
\begin{align}
    1=\left[1+f-u_r\vert_{\xi=1+f(\theta,\phi)}\right]^2+\left[u_\theta\vert_{\xi=1+f(\theta,\phi)}\right]^2+\left[u_\phi\vert_{\xi=1+f(\theta,\phi)}\right]^2 .\label{eq:deform_disp_relation}
\end{align}
Eqn (\ref{eq:deform_disp_relation}) indicates that, for any material point on the deformed surface, the magnitude of the difference between its current position and the displacement it experienced is equal to one. In the weakly deformable limit ($\alpha\ll 1$), the shape of the particle deviates slightly from that of a sphere. We expand all the variables in the problem in the form of regular asymptotic expansions and use the domain perturbation method.\citep{rangasbook,garyleal} The corresponding modified boundary conditions are obtained at different orders in $\alpha$. The series solutions to the Stokes and elasticity equations are used to obtain the variables that satisfy the modified boundary conditions. In the following discussion, we represent the quantities using either boldface or index notation, choosing whichever form is simpler for a given expression.

In general, the ambient velocity field at an arbitrary point relative to the particle centre [$\mathbf{v^{\infty}(\mathbf{x})}$] is obtained by expanding the field about the centre of the particle using a Taylor series as given by
\begin{align}
  \mathbf{v^{\infty}(\mathbf{x})}= \mathbf{v^{\infty}}(\mathbf{x}=0) +\mathbf{\mathbf{x} \cdot \nabla v^{\infty}}(\mathbf{x}=0)+\frac{1}{2}\left[\mathbf{xx}:\mathbf{\nabla \nabla} \mathbf{v^{\infty}}(\mathbf{x}=0)\right]+...\,.\label{eq:general ambient field expansion about particle centre}
\end{align}
The first, second, and third terms on the right-hand side of eqn (\ref{eq:general ambient field expansion about particle centre}) are the uniform, linear, and quadratic components of the ambient flow, respectively. In the present work, we consider the quadratic component as the ambient flow ($\mathbf{v}_{q}$) for the particle. We refer to the unbounded ambient flow ($\mathbf{v}_{q}$) as the general quadratic flow. The flow can be expressed in terms of a third-rank tensor $K_{ijk}$\citep{nadim1991motion} and is given by
\begin{align}
&(v_{q})_k=x_ix_j K_{ijk}\,, \label{eq:velocity in general quadratic}\\
 &K_{ijk}= \frac{1}{2}\nabla_i \nabla_j v^{\infty}_k (\mathbf{x}=0)\,.\label{eq:Kijk expression}
\end{align}
In eqn (\ref{eq:velocity in general quadratic}) and (\ref{eq:Kijk expression}), $K_{ijk}$ is symmetric with respect to the first two indices. Further, $K_{ijk}$ can be expressed in terms of irreducible components that are first-, second-, and third-order tensors represented by $\boldsymbol{\tau}$, $\mathbf{Q}$ and $\boldsymbol{\gamma}$, respectively.\citep{coope1965irreducible,nadim1991motion} The tensors are given by
\begin{align}
  &\tau_{l}= K_{ppl}\,, \label{eq:tau expression}\\
  &Q_{lm}= K_{lpq}\epsilon_{qpm}+K_{mpq}\epsilon_{qpl}\, \,\text{and}\\
  &\gamma_{ijk}= \frac{1}{3} \left(K_{ijk}+K_{kij}+K_{jki}\right)-\frac{1}{15}\left(\delta_{ij}K_{ppk}+\delta_{ik}K_{ppj}+\delta_{jk}K_{ppi}\right)\,. \label{eq:gamma expression}
\end{align}
The components $\mathbf{Q}$ and $\boldsymbol{\gamma}$ are symmetric and traceless with respect to any pair of their indices. The velocity and pressure fields of the general quadratic flow can be expressed in terms of the irreducible components as
\begin{equation}
  (v_{q})_k= x_i x_j \gamma_{ijk}-\frac{1}{3}(\epsilon_{kij}x_l Q_{li} x_j) + \frac{1}{5}(2\xi^2 \delta_{kj}-x_k x_j)\tau_j
\end{equation}
and
\begin{equation}
    p_q =P_0+2(\tau_i x_i)\,,\label{eq:ambient pressure of general quadratic flow}
\end{equation}
respectively. Here, $P_0$ is the reference pressure at the centre of the undeformed particle. In our analysis, we assume the flow component $\boldsymbol{\tau}$ and the velocity of the particle are both aligned with the $z$-axis. The assumption is justified by the fact that the $\boldsymbol{\tau}$, corresponding to the quadratic component of the Poiseuille flow, is aligned with the channel length ($z$-axis). In addition, the general quadratic flow exerts a hydrodynamic force at the leading-order along $\boldsymbol{\tau}$ and the particle is constrained to move with a fixed velocity in the same direction. Hence, both $\boldsymbol{\tau}$ and the particle velocity are taken along the $z$-axis. We discuss the series solutions to the governing equations in Section \ref{subsec:series solution to Stokes and elasticity eq} and the domain perturbation method in Section \ref{subsec:domain perturbation method}. We provide the solution procedure to obtain the variables ($\mathbf{v}$, $ \boldsymbol{\sigma}$, $p$, $\mathbf{u}$, $f$, $ \hat{\boldsymbol{\sigma}}$, $\mathbf{F}$, $\mathbf{T}$) in Section \ref{subsec:solution procedure}.

\subsection {Series solutions to the Stokes and Navier elasticity equations}\label{subsec:series solution to Stokes and elasticity eq}
The series solutions in the spherical coordinate system to eqn (\ref{eq:stokes}) and (\ref{eq:continuity}) are given in Appendix A, while the solution to eqn (\ref{eq:elasticity}) is given in Appendix B. In the solution given in eqn (\ref{eq:ur series})-(\ref{eq:uphi series}), the point force and the point torque are denoted by $F_{i}$ and $T_{i}$ (with $i=x,y,z$ being the Cartesian components), respectively. The displacement field corresponding to the point force and the point torque decays as $1/\xi$ and $1/\xi^2$, respectively [for details, see Appendix A of Verma et al.\citep{verma2025dynamics}]. We analyse the problem in a particle-fixed coordinate system whose origin coincides with the centre of the undeformed particle, and the point torque inhibits the rotation of the particle. The constants that correspond to pure translation ($c1[0,1]$, $c1[1,1]$, and $c0[1,1]$) and pure rotation with no deformation ($a1[0,1]$, $a1[1,1]$, and $a0[1,1]$) in the series solution given in eqn (\ref{eq:ur series})-(\ref{eq:uphi series}) are, therefore, set to zero.
\subsection{Domain perturbation method }\label{subsec:domain perturbation method}
In this section, we discuss the domain perturbation method, following the approach outlined in Verma et al..\citep{verma2025dynamics} The method is used to transfer the boundary conditions from the deformed surface of the particle to its undeformed surface.\citep{rangasbook,garyleal} The surface deformation $f$ defined in eqn (\ref{eq:surface}) must be determined as a part of the solution. The velocity boundary condition (no-slip and no-penetration) on the deformed surface is given by
\begin{align}
  \mathbf{v}=0\,,
  \label{eq:velocity bc}
\end{align}
where $\mathbf{v}=\mathbf{v}_d+\mathbf{v}_{q}-1\mathbf{\hat{k}}$. The disturbance velocity and the general quadratic flow fields are denoted by $\mathbf{v}_d$ and $\mathbf{v}_{q}$, respectively. The stress continuity on the deformed surface is given by
\begin{align}
  \hat{\boldsymbol{\sigma}}\cdot\mathbf{\hat{n}}=\alpha \, \boldsymbol{\sigma}\cdot\mathbf{\hat{n}}\,. \label{eq:stress bc}
\end{align}
Here, $\mathbf{\hat{n}}$ is the outward unit normal from the deformed surface, defined by $\mathbf{\hat{n}}= (\nabla S/|\nabla S|)$, where $S= \xi-1-f(\theta,\phi)$. 
The stress boundary conditions in terms of different stress components are given by
\begin{align}
  \hat{\sigma}_{rr}-\frac{\hat{\sigma}_{r\theta}}{\xi}\frac{\partial{f}}{\partial{\theta}}-\frac{\hat{\sigma}_{r\phi}}{\xi \sin\theta}\frac{\partial {f}}{\partial{\phi}}=\alpha\left[\sigma_{rr}-\frac{\sigma_{r\theta}}{\xi}\frac{\partial{f}}{\partial{\theta}}-\frac{\sigma_{r\phi}}{\xi \sin\theta}\frac{\partial {f}}{\partial{\phi}}\right]\,,\label{eq:rr stress bc}
\end{align}
\begin{align}
  \hat{\sigma}_{r\theta}-\frac{\hat{\sigma}_{\theta\theta}}{\xi}\frac{\partial{f}}{\partial{\theta}}-\frac{\hat{\sigma}_{\theta\phi}}{\xi \sin\theta}\frac{\partial {f}}{\partial{\phi}}=\alpha\left[\sigma_{r\theta}-\frac{\sigma_{\theta\theta}}{\xi}\frac{\partial{f}}{\partial{\theta}}-\frac{\sigma_{\theta \phi}}{\xi \sin\theta}\frac{\partial {f}}{\partial{\phi}}\right]\label{eq:rtheta stress bc}
\end{align}
and
\begin{align}
  \hat{\sigma}_{r\phi}-\frac{\hat{\sigma}_{\theta\phi}}{\xi}\frac{\partial{f}}{\partial{\theta}}-\frac{\hat{\sigma}_{\phi\phi}}{\xi \sin\theta}\frac{\partial {f}}{\partial{\phi}}=\alpha\left[\sigma_{r\phi}-\frac{\sigma_{\theta\phi}}{\xi}\frac{\partial{f}}{\partial{\theta}}-\frac{\sigma_{\phi \phi}}{\xi \sin\theta}\frac{\partial {f}}{\partial{\phi}}\right]\,.\label{eq:rphi stress bc}
\end{align}
Here, the $\hat{\sigma}_{ij}$ and $\sigma_{ij}$ with $i,j=r,\theta,\phi$ are the components of the solid and fluid stress tensors, respectively. In the domain perturbation method, we first performed a Taylor series expansion of the boundary conditions, defined in eqn (\ref{eq:velocity bc})-(\ref{eq:rphi stress bc}) about $\xi=1$, and then the regular asymptotic expansions of the variables in the problem are substituted into these expanded boundary conditions. The regular expansions of the variables as series in $\alpha$ are given by
\begin{equation}
 \mathbf{v} = \mathbf{v}^{(0)}+\alpha\mathbf{v}^{(1)}+\alpha^2\mathbf{v}^{(2)}+...\, , \label{eq:vel_exp_in_alpha}
\end{equation}
\begin{equation}
 \boldsymbol{\sigma} =\boldsymbol{\sigma}^{(0)}+\alpha\boldsymbol{\sigma}^{(1)}+\alpha^2\boldsymbol{\sigma}^{(2)}+...\, ,\label{eq:fluidstress_exp_in_alpha}
\end{equation}
\begin{equation}
 p =p^{(0)}+\alpha p^{(1)}+\alpha^2 p^{(2)}+...\, ,
\end{equation}
\begin{equation}
 \mathbf{u} =\alpha\mathbf{u}^{(1)}+\alpha^2\mathbf{u}^{(2)}+...\, ,\label{eq:displacement_exp_in_alpha}
\end{equation}
\begin{equation}
f =\alpha f^{(1)}+\alpha^2 f^{(2)}+...\, ,\label{eq:deform_exp_in_alpha}
\end{equation}
\begin{equation}
 \hat{\boldsymbol{\sigma}} =\alpha\hat{\boldsymbol{\sigma}}^{(1)}+\alpha^2\hat{\boldsymbol{\sigma}}^{(2)}+...\, ,\label{eq:solidstress_exp_in_alpha}
\end{equation}
\begin{equation}
 \mathbf{F} =\mathbf{F}^{(0)}+\alpha\mathbf{F}^{(1)}+\alpha^2\mathbf{F}^{(2)}+...
\end{equation}
and
\begin{equation}
 \mathbf{T} =\mathbf{T}^{(0)}+\alpha\mathbf{T}^{(1)}+\alpha^2\mathbf{T}^{(2)}+...\, .\label{eq:torque_exp_in_alpha}
\end{equation}  
It is important to note that the leading-order terms in the series of the displacement field  [eqn (\ref{eq:displacement_exp_in_alpha})], surface deformation [eqn (\ref{eq:deform_exp_in_alpha})], and solid stress field [eqn (\ref{eq:solidstress_exp_in_alpha})] arising from the deformation appear at \textit{O}($\alpha$). 
\subsubsection{Modified velocity boundary conditions on the undeformed surface ($\xi=1$)}\label{subsubsec:mod vel bc}
The Taylor series expansion of eqn (\ref{eq:velocity bc}) about the undeformed surface ($\xi=1$) is carried out, and the expansion of velocity and surface deformation from eqn (\ref{eq:vel_exp_in_alpha}) and (\ref{eq:deform_exp_in_alpha}), respectively, are substituted to obtain the modified velocity boundary conditions at successive orders in $\alpha$. At leading-order, the boundary conditions are given by
\begin{align}
  \mathbf{v}^{(0)}&=0 \,\,  \mbox{as $\xi$ =1}\,, \label{eq:vel_v0_bc1} \\
  &\rightarrow \mathbf{v}_q -1  \mathbf{\hat{k}} \,\, \mbox{as $\xi \rightarrow \infty$} \,.  
  \label{eq:vel_v0_bc2}
\end{align}
At \textit{O}($\alpha$), it is
\begin{equation} \mathbf{v}^{(1)}+f^{(1)}\frac{\partial{\mathbf{v}^{(0)}}}{\partial{\xi}}=0   \label{eq:vel_v1_bc}
\end{equation}
and at \textit{O}($\alpha^2$), it is
\begin{equation} \mathbf{v}^{(2)}+f^{(1)}\frac{\partial{\mathbf{v}^{(1)}}}{\partial{\xi}}+f^{(2)}\frac{\partial{\mathbf{v}^{(0)}}}{\partial{\xi}}+\frac{\left[f^{(1)}\right]^2}{2}\frac{\partial^2{\mathbf{v}^{(0)}}}{\partial{\xi}^2}=0 \,.\label{eq:vel_v2_bc}
\end{equation}
The velocity fields $\mathbf{v}^{(1)}$ and $\mathbf{v}^{(2)}$  are the disturbance fields and decay as $\xi \rightarrow \infty$. Note that the boundary conditions in eqn (\ref{eq:vel_v1_bc}) and (\ref{eq:vel_v2_bc}) are nonlinear. Although the governing equations are linear, the nonlinearity stems from the fact that the unmodified boundary conditions are applied on the deformed surface, whose shape is a part of the solution.
\subsubsection{Modified stress boundary conditions on the undeformed surface ($\xi=1$)}\label{subsubsec:mod stress bc}
The Taylor series expansion of eqn (\ref{eq:stress bc}) about the undeformed surface ($\xi=1$) is carried out, and the expansions of fluid stress, surface deformation, and solid stress from eqn (\ref{eq:fluidstress_exp_in_alpha}), (\ref{eq:deform_exp_in_alpha}), and (\ref{eq:solidstress_exp_in_alpha}), respectively, are substituted to obtain the modified stress boundary conditions at various orders in $\alpha$. At \textit{O}($\alpha$), the stress boundary conditions are
\begin{equation}
  \hat{\sigma}^{(1)}_{rr} = \sigma^{(0)}_{rr}\,, \label{eq:Oalpha_stressbc_r}
\end{equation}
\begin{equation}
  \hat{\sigma}^{(1)}_{r\theta} = \sigma^{(0)}_{r\theta} \,,\label{eq:Oalpha_stressbc_theta}
\end{equation}
\begin{equation}
  \hat{\sigma}^{(1)}_{r\phi} = \sigma^{(0)}_{r\phi}\label{eq:Oalpha_stressbc_phi}
\end{equation}
and at \textit{O}($\alpha^2$), they are
\begin{equation}
    \hat{\sigma}_{rr}^{(2)}=\sigma_{rr}^{(1)}+f^{(1)} \frac{\partial}{\partial{\xi}}\left[\sigma_{rr}^{(0)}-\hat{\sigma}_{rr}^{(1)}\right]\,,\label{eq:Oalpha^2_stressbc_r}
\end{equation}
\begin{align}
    \hat{\sigma}_{r\theta}^{(2)}=&\sigma_{r\theta}^{(1)}-f^{(1)} \frac{\partial}{\partial{\xi}}\left[\hat{\sigma}_{r\theta}^{(1)}-\sigma_{r\theta}^{(0)}\right]+\frac{\partial{f^{(1)}}}{\partial{\theta}}\left[\hat{\sigma}_{\theta \theta}^{(1)}-\sigma_{\theta \theta}^{(0)}\right] \nonumber\\&+\frac{1}{\sin\theta}\frac{\partial{f^{(1)}}}{\partial{\phi}}\left[\hat{\sigma}_{\theta \phi}^{(1)}-\sigma_{\theta \phi}^{(0)}\right]\,,\label{eq:Oalpha^2_stressbc_theta}
\end{align}
\begin{align}
    \hat{\sigma}_{r\phi}^{(2)}=&\sigma_{r\phi}^{(1)}-f^{(1)} \frac{\partial}{\partial{\xi}}\left[\hat{\sigma}_{r\phi}^{(1)}-\sigma_{r\phi}^{(0)}\right]+\frac{\partial{f^{(1)}}}{\partial{\theta}}\left[\hat{\sigma}_{\theta \phi}^{(1)}-\sigma_{\theta \phi}^{(0)}\right] \nonumber\\&+ \frac{1}{\sin\theta}\frac{\partial{f^{(1)}}}{\partial{\phi}}\left[\hat{\sigma}_{\phi \phi}^{(1)}-\sigma_{\phi \phi}^{(0)}\right]\,.\label{eq:Oalpha^2_stressbc_phi}
\end{align}

The surface deformation at different orders in $\alpha$, such as $f^{(1)}$ and $f^{(2)}$ appearing in the modified boundary conditions are obtained by expanding eqn (\ref{eq:deform_disp_relation}) using the Taylor series about $\xi=1$, followed by substituting eqn (\ref{eq:displacement_exp_in_alpha}) and (\ref{eq:deform_exp_in_alpha}). At \textit{O}($\alpha$), $f^{(1)}$ is given by
\begin{equation}
    f^{(1)}=\left.u_r^{(1)}\right \vert_{\xi=1}\,. \label{eq:f1_Expression}
\end{equation}
At \textit{O}($\alpha^2$), $f^{(2)}$ is given by
\begin{align}
    f^{(2)}=\left(u_r^{(2)} + f^{(1)} \frac{\partial{u_r^{(1)}}}{\partial{\xi}} -\frac{\left[u_\theta^{(1)}\right]^2 + \left[u_\phi^{(1)}\right]^2}{2}\right)\left. \vphantom{\frac{\left[u_\theta^{(1)}\right]^2}{2}} \right\vert_{\xi=1}\,. \label{eq:f2_Expression}
\end{align}
The modified boundary conditions derived in this section, along with the series solutions described in the previous sections, are used to calculate all the variables defined in eqn (\ref{eq:vel_exp_in_alpha})-(\ref{eq:torque_exp_in_alpha}).

\subsection{Solution procedure}\label{subsec:solution procedure}
Here, we describe the steps involved in calculating the velocity, pressure, displacement, point force, and point torque until  \textit{O}($\alpha^2$), as the particle translates in the general quadratic flow.
\begin{enumerate}
    \item []
    \item{At \textit{O}(1), the constants in Appendix A, eqn (\ref{eq:vr series})-(\ref{eq:vphi series}) are determined by applying the velocity boundary conditions given in eqn (\ref{eq:vel_v0_bc1}) and (\ref{eq:vel_v0_bc2}), and corresponding $p^{(0)}$ and $\boldsymbol{\sigma}^{(0)}$ are calculated.}
    \item[]
    \item{Using stress continuity at \textit{O}($\alpha$) given in eqn (\ref{eq:Oalpha_stressbc_r})-(\ref{eq:Oalpha_stressbc_phi}), the constants in Appendix B that appear in $\boldsymbol{\hat{\sigma}}^{(1)}$, $\mathbf{F}^{(0)}$ and $\mathbf{T}^{(0)}$ are determined. Subsequently, $\mathbf{u}^{(1)}$ and $f^{(1)}$ are calculated.}
    \item[]
    \item{At \textit{O}($\alpha$), the constants in Appendix A, eqns. (\ref{eq:vr series})–(\ref{eq:vphi series}), are determined using the velocity boundary condition given in eqn (\ref{eq:vel_v1_bc}) and in turn calculate $p^{(1)}$ and $\boldsymbol{\sigma}^{(1)}$.}
    \item[]   
    \item{Using stress continuity at \textit{O}($\alpha^2$) given in eqn (\ref{eq:Oalpha^2_stressbc_r})-(\ref{eq:Oalpha^2_stressbc_phi}), the constants in Appendix B that appear in $\boldsymbol{\hat{\sigma}}^{(2)}$, $\mathbf{F}^{(1)}$ and $\mathbf{T}^{(1)}$ are obtained. Subsequently, $\mathbf{u}^{(2)}$ and $f^{(2)}$ are calculated.}
    \item[]   
    \item{At \textit{O}($\alpha^2$), the constants in Appendix A, eqn (\ref{eq:vr series})-(\ref{eq:vphi series}) are determined by applying the velocity boundary condition given in eqn (\ref{eq:vel_v2_bc}) and corresponding $p^{(2)}$ and $\boldsymbol{\sigma}^{(2)}$ are calculated. The point force (torque) at \textit{O}($\alpha^2$) is obtained directly using the fluid stress (moment of fluid stress about origin) at the same order, in the absence of body force as discussed in Appendix C. When the body force (torque) is present, additional terms corresponding to the body force (torque) must be considered. Alternatively, the next-order stress boundary conditions [\textit{O}($\alpha^3$)] can be applied to determine the point force and the point torque.}
\end{enumerate}
In the next section, we discuss the results of our analysis.
\section{Results}\label{sec:results}
First, we analysed the dynamics of the particle and obtained the results in a general quadratic flow. The results are then simplified for the quadratic component of all three Poiseuille flows, as shown in Fig. \ref{fig:schematic_pp_hp_flow}. The confinement ratios in each of the flows are defined based on their geometry. In the elliptical Poiseuille flow, the confinement ratios along the $x$ and $y$ axes are defined as $\psi_x=R_0/a$ and $\psi_y=R_0/b$, respectively.  For the plane Poiseuille and Hagen-Poiseuille flows, the confinement ratios are defined as $\psi=R_0/h$ and $\psi=R_0/R_t$, respectively. We considered the confinement ratio to be much less than one, assuming the channel size is much larger than the particle size. We first discuss the expressions of the variables ($\mathbf{v}$, $p$, $\mathbf{u}$, $f$, $\mathbf{F}$ and $\mathbf{T}$) and then plot both the streamlines and the deformed shape of the particle in section \ref{subsec:Analyses of the velocity and deformation}.

Here, we present the results for the particle translating in a general quadratic flow. The components of the disturbance velocity and the pressure fields in the fluid at \textit{O}(1) and the components of the displacement field at \textit{O}($\alpha$) are given in the ESI$^{\dag}$ (section 1). At \textit{O}($\alpha$), the surface deformation is obtained as
\begin{align}
    {f^{(1)}_{q}}=&\frac{1}{256 (2 + \Gamma) (2 + 3 \Gamma)}\left[\vphantom{\frac{1}{256 (2 + \Gamma) (2 + 3 \Gamma)}}6 \cos\theta(\Gamma^2 (128 + 105 \gamma_{333}) + 4 (80 + 35 \gamma_{333} \right.\nonumber\\&\left.- 48 \tau) + 8 \Gamma (64 + 35 \gamma_{333} - 16 \tau) - 140 (2 + \Gamma)(2 + 3 \Gamma)(2 \gamma_{223} \right.\nonumber\\&\left.+ \gamma_{333}) \cos2 \phi \sin^2 \theta)+(2+\Gamma)(350 (2 + 3 \Gamma) \gamma_{333} \cos3 \theta + 35 (2 \right.\nonumber\\&\left.+ 3 \Gamma)[ -4 (4 \gamma_{122} + \gamma_{133}) \cos3 \phi \sin^3 \theta + 3 \gamma_{233} (\sin \theta + 5 \sin3 \theta) \right.\nonumber\\&\left.\times\sin \phi + 3 \cos \phi (\gamma_{133} \sin \theta + 5 \gamma_{133} \sin3 \theta + 32 \gamma_{123} \cos \theta \sin^2 \theta \right.\nonumber\\&\left.\times\sin \phi) - 4 (4 \gamma_{222} + 3 \gamma_{233}) \sin^3 \theta \sin3 \phi ])\vphantom{\vphantom{\frac{1}{256 (2 + \Gamma) (2 + 3 \Gamma)}}}\right] -\frac{P_0}{2+3\Gamma} \,.
\end{align}
Here, $\Gamma=\lambda/G$ as defined following eqn (\ref{eq:nondimensional variables}). Note that the deformation associated with $P_0$ represents purely radial compression on account of the compressible nature of the particle. The expressions for the components of disturbance velocity ($\mathbf{v}^{(1)}_{d_q}$) and the pressure ($p^{(1)}_{d_q}$) fields at \textit{O}($\alpha$), and the components of displacement ($\mathbf{u}^{(2)}_{q}$), velocity ($\mathbf{v}^{(2)}_{d_q}$), and the pressure ($p^{(2)}_{d_q}$) fields at \textit{O}($\alpha^2$) are lengthy and omitted here for brevity. The components of $\mathbf{v}^{(1)}_{d_q}$ and the $p^{(1)}_{d_q}$ are obtained by substituting the corresponding constants ($a[m,n]$, $\tilde{a}[m,n]$, $b[m,n]$, $\tilde{b}[m,n]$, $v[m,n]$, and $\tilde{v}[m,n]$) (ESI$^{\ddag}$) in the velocity [eqn (\ref{eq:vr series})-(\ref{eq:vphi series})] and pressure series [eqn (\ref{eq:press series})], respectively. The components of $\mathbf{u}^{(2)}_{q}$ are obtained by substituting the corresponding constants ($a1[m,n]$, $a0[m,n]$, $b1[m,n]$, $b0[m,n]$, $c1[m,n]$, and $c0[m,n]$) (ESI$^{\ddag}$), in the displacement series [eqn (\ref{eq:ur series})-(\ref{eq:uphi series})]. Further, the surface deformation at \textit{O}($\alpha^2$) is obtained using eqn (\ref{eq:f2_Expression}). At \textit{O}($\alpha^2$), the components of the $\mathbf{v}^{(2)}_{d_q}$ and the $p^{(2)}_{d_q}$ are obtained by substituting the corresponding constants ($a[m,n]$, $\tilde{a}[m,n]$, $b[m,n]$, $\tilde{b}[m,n]$, $v[m,n]$, and $\tilde{v}[m,n]$) (ESI$^{\ddag}$), in the velocity series [eqn (\ref{eq:vr series})-(\ref{eq:vphi series})] and pressure series [eqn (\ref{eq:press series})], respectively.
The point force until \textit{O}($\alpha^2$) in dimensional form is obtained as
\begin{align}
    \mathbf{F}_{q}=&6\pi\mu V_0 R_{0}\left(\left[1-\frac{\tau}{3}\right]\mathbf{\hat{k}}+\alpha\left[\frac{P_0(\tau-1)}{2+3\Gamma}\right]\mathbf{\hat{k}}+ \alpha^2\frac{2}{3} \left[\vphantom{\frac{2}{3}}a[1,1]\mathbf{\hat{i}}\right.\right.\nonumber\\&\left.\left.+ \tilde{a}[1,1]\mathbf{\hat{j}}+a[0,1]\mathbf{\hat{k}}\vphantom{\frac{2}{3}}\right]\right)\,. \label{eq:force in general quadratic}
\end{align}
The constants $a[1,1]$, $\tilde{a}[1,1]$, and $a[0,1]$ (ESI$^{\ddag}$) are among the constants determined from the velocity boundary conditions at \textit{O}($\alpha^2$). The leading-order point force in eqn (\ref{eq:force in general quadratic}) reduces to the Stokes drag on a rigid sphere translating along the $\boldsymbol{\tau}$ ($z$-axis) in the general quadratic flow. Due to the elastic effects, the point force at \textit{O}($\alpha$) acts along the $z$-axis and is proportional to the reference pressure at the centre of the undeformed sphere [defined following eqn (\ref{eq:ambient pressure of general quadratic flow})]. It is important to note that the force at \textit{O}($\alpha^2$) has components along all three axes in the general quadratic flow. However, as we see in later sections, the force acts only along the $z$-axis in the elliptical Poiseuille, plane Poiseuille, and Hagen-Poiseuille flows. The drag force at the leading-order scales as $\mu V_0 R_0$, at $O(\alpha)$ as $\mu^2 V_0^2/G$, and at $O(\alpha^2)$ as $(\mu^3 V_0^3/G^2)(1/R_0)$. The point torque until \textit{O}($\alpha^2$) in dimensional form is obtained as
\begin{align}
    \mathbf{T}_{q}=& 8\pi\mu V_0 R_{0}^2\left(\alpha\left[Z_{1}\mathbf{\hat{i}} +Z_{2}\mathbf{\hat{j}}+Z_{3}\mathbf{\hat{k}}\right]+\alpha^2\left[v[1,1]\mathbf{\hat{i}} +\tilde{v}[1,1]\mathbf{\hat{j}}\right.\right.\nonumber\\&\left.\left.+v[0,1]\mathbf{\hat{k}} \vphantom{\alpha^2}\right]\right)\,. \label{eq:torque in general quadratic}
\end{align}
The expressions for $Z_1$, $Z_2$, and $Z_3$ are given in the ESI$^\dag$ (section 1). The constants $v[1,1]$, $\tilde{v}[1,1]$, and $v[0,1]$ (ESI$^{\ddag}$) are among the constants determined from the velocity boundary conditions at \textit{O}($\alpha^2$). The leading-order point torque in eqn (\ref{eq:torque in general quadratic}) appears at \textit{O}($\alpha$) and exhibits a scaling of $(\mu^2 V_0^2/G)(R_0)$, while the torque at \textit{O}($\alpha^2$) exhibits a scaling of ($\mu^3 V_0^3/G^2$). The components of the torque at \textit{O}($\alpha$) and \textit{O}($\alpha^2$) act along all three axes. However, the torque is found to be zero in all three Poiseuille flows (Fig. \ref{fig:schematic_pp_hp_flow}), consistent with the fact that a particle located at the centreline of the channel does not experience a hydrodynamic torque. We present the simplification of quadratic results in sections \ref{subsec:Simplification for the elliptical Poiseuille flow}, \ref{subsec:Simplification for the plane Poiseuille flow}, and \ref{subsec:Simplification for the Hagen Poiseuille flow}. In Poiseuille flows, we constrain the particle to translate with the centreline velocity ($V_{max}$).

\subsection{Simplification for the elliptical Poiseuille flow}\label{subsec:Simplification for the elliptical Poiseuille flow}
In this section, we present the results for the particle translating along the centreline in the quadratic component of the elliptical Poiseuille flow. The undisturbed velocity field is given by
\begin{align}
    \mathbf{v}_{e}=\left[1-\left(\frac{x}{a}\right)^2-\left(\frac{y}{b}\right)^2\right]\mathbf{\hat{k}}\,,
\end{align}
The quadratic component of the elliptical Poiseuille flow ($\mathbf{v}_{q_e}$) is represented in terms of third rank tensor $\mathbf{K}_{q_e}$ as given by 
\begin{align}
    \mathbf{v}_{q_e}=&\mathbf{x}\mathbf{x}:\mathbf{K}_{q_e}\,,\\
    \mathbf{K}_{q_e}=&-\left[\, \psi_x^2\,  \mathbf{\hat{i}}\, \mathbf{\hat{i}}\, \mathbf{\hat{k}}+\psi_y^2\,  \mathbf{\hat{j}}\, \mathbf{\hat{j}}\, \mathbf{\hat{k}}\right]\,.\label{eq:ambient flow in Kijk_pa}
\end{align}
The irreducible components ($\boldsymbol{\tau}_{q_e}$, $\mathbf{Q}_{q_e}$, and $ \boldsymbol{\gamma}_{q_e}$) are obtained by substituting eqn (\ref{eq:ambient flow in Kijk_pa}) in eqn (\ref{eq:tau expression})-(\ref{eq:gamma expression}). The corresponding components of the undisturbed velocity and the pressure fields are given by
\begin{align}
    &{v_{q_e,r}}=-\frac{1}{2 }\left[\vphantom{\frac{1}{2 }}\xi^2 \cos\theta (\psi_x^2 + \psi_y^2 + (\psi_x^2 - \psi_y^2) \cos 2\phi) \sin^2\theta\right]\,,\label{eq:ambient vel_r_pa} \\ 
    &v_{q_e,\theta}=\frac{1}{2 }\left[\vphantom{\frac{1}{2 }}\xi^2 (\psi_x^2 + \psi_y^2 + (\psi_x^2 - \psi_y^2) \cos 2\phi) \sin^3\theta\right]\,,\\
    &v_{q_e,\phi}=0\,\,\text{and }\\ 
    &p_{q_e}=P_0-2 \xi (\psi_x^2 + \psi_y^2) \cos\theta \,,\label{eq:ambient pressure_pa}
\end{align}
respectively. The ambient field gets disturbed by the translation of the particle. The components of the disturbance velocity ($\mathbf{v}^{(0)}_{d_{q_e}}$) and the pressure ($p^{(0)}_{d_{q_e}}$) fields in the fluid at the leading-order and the components of the displacement ($\mathbf{u}^{(1)}_{{q_e}}$) field at \textit{O}($\alpha$) are given in the ESI$^{\dag}$ (section 2). The surface deformation at \textit{O}($\alpha$) is obtained as
\begin{align}
f^{(1)}_{q_e}=& \frac{1}{8}\left[\vphantom{\frac{1}{8}}8 (\mathcal{A}_1 + \mathcal{E}_1) - 5 \mathcal{A}_3 \cos 2\theta - 5 \mathcal{A}_4 \cos 2\phi \sin^2 \theta\right]\cos\theta \nonumber\\&-\frac{P_0}{2+3\Gamma}\,.\label{eq:order alpha deformation expression_pa}
\end{align}
The expressions for $\mathcal{A}_1$, $\mathcal{A}_3$, $\mathcal{A}_4$ and $\mathcal{E}_1$ are given in the ESI$^{\dag}$ (section 2). The components of disturbance velocity ($\mathbf{v}^{(1)}_{d_{q_e}}$) and the pressure ($p^{(1)}_{d_{q_e}}$) fields in the fluid at \textit{O}($\alpha$) and the displacement ($\mathbf{u}^{(2)}_{{q_e}}$) field at \textit{O}($\alpha^2$) are given in the ESI$^{\dag}$ (section 2).
At \textit{O}($\alpha^2$), the surface deformation ($f^{(2)}_{q_e}$) is obtained as
\begin{align}
f^{(2)}_{q_e}=&\mathcal{L}_1 - \mathcal{L}_2 \cos2\theta + \mathcal{L}_3 \cos4\theta + \mathcal{L}_4 \cos6\theta - (\mathcal{L}_5 + \mathcal{L}_6 \cos2\theta \nonumber\\&+ \mathcal{L}_7 \cos4\theta) \cos2\phi \sin^2\theta + (\mathcal{L}_8 + \mathcal{L}_9 \cos2\theta) \cos4\phi \sin^4\theta \nonumber\\&+ P_0^2\left[\mathcal{L}_{10}\right]+P_0\left[\mathcal{L}_{11} \cos 3\theta + \cos \theta ( \mathcal{L}_{12} + \mathcal{L}_{13} \cos 2\phi \sin^2 \theta )\right] \,.\label{eq:order^2 alpha deformation expression_pa}
\end{align}
The expressions for $\mathcal{L}_n$ (with $n$ =1,2,...,13) are given in the ESI$^{\dag}$ (section 2). Note that the surface deformations $f^{(1)}_{q_e}$ and $f^{(2)}_{q_e}$ are functions of the azimuthal angle ($\phi$) due to the asymmetry of the channel geometry about the $z$-axis. The components of $\mathbf{v}^{(2)}_{d_{q_e}}$ and $p^{(2)}_{d_{q_e}}$ for the elliptical Poiseuille flow are obtained by substituting the irreducible components $\boldsymbol{\gamma}_{q_e}$, $\mathbf{Q}_{q_e}$, and $\boldsymbol{\tau}_{q_e}$ into the corresponding results ($\mathbf{v}^{(2)}_{d_{q}}$ and $p^{(2)}_{d_{q}}$) of the general quadratic flow. The point force until \textit{O}($\alpha^2$) in dimensional form is obtained as
\begin{align}
    \mathbf{F}_{q_e}=&6\pi\mu V_0 R_{0}\left(\left[1+\frac{(\psi_x^2+\psi_y^2)}{3}\right]-\alpha P_0\left[\frac{1+\psi_x^2+\psi_y^2}{(2+3\Gamma)}\right]\right.\nonumber\\& \left.+\alpha^2 \left[ \vphantom{\frac{1 + 2 (\psi_x^2 + \psi_y^2)}{ (2 + 3 \Gamma)^2}}\mathcal{X}_1 + \mathcal{X}_2 \psi_x^2 + \mathcal{X}_3 \psi_x^4 + \mathcal{X}_4 \psi_x^6 + (\mathcal{X}_2 + \mathcal{X}_6 \psi_x^2 \right.\right. \nonumber\\& \left.\left.+ \mathcal{X}_5 \psi_x^4) \psi_y^2 + (\mathcal{X}_3 + \mathcal{X}_5 \psi_x^2) \psi_y^4 + \mathcal{X}_4 \psi_y^6 \right.\right. \nonumber\\& \left.\left.+P_0^2\left(\frac{1 + 2 (\psi_x^2 + \psi_y^2)}{ (2 + 3 \Gamma)^2}\right)\vphantom{\frac{1 + 2 (\psi_x^2 + \psi_y^2)}{ (2 + 3 \Gamma)^2}}\right]\right)\mathbf{\hat{k}} \,.\label{eq:force in ep flow}
\end{align}
The expressions for $\mathcal{X}_n$ (with $n=1,2,...,6$) are given in the ESI$^{\dag}$ (section 2). The effects of elasticity in eqn (\ref{eq:force in ep flow}) appear at \textit{O}($\alpha$) and \textit{O}($\alpha^2$). Additionally, the point force is aligned with the velocity of the particle ($z$-axis) and balances the hydrodynamic drag. The results of the elliptical Poiseuille flow are further simplified for the plane Poiseuille and Hagen-Poiseuille flows in sections \ref{subsec:Simplification for the plane Poiseuille flow} and \ref{subsec:Simplification for the Hagen Poiseuille flow}, respectively.
\subsection{Simplification for the plane Poiseuille flow}\label{subsec:Simplification for the plane Poiseuille flow}
The quadratic flow, along with the disturbance velocity, pressure, displacement, deformation, and point force at different orders in $\alpha$ are obtained for the plane Poiseuille flow by substituting $\psi_x=\psi$ and $\psi_y=0$ in the corresponding expressions for the elliptical Poiseuille flow. The components of undisturbed velocity and the pressure fields for the quadratic component of the plane Poiseuille flow are obtained as
\begin{align}
    &{v_{q_p,r}}=-\xi^2\psi^2  \cos\theta \cos^2\phi \sin^2\theta\,,\\
    &v_{q_p,\theta}=\xi^2\psi^2  \cos^2\phi \sin^3\theta\,,\\
    &v_{q_p,\phi}=0\,\,\text{and }\\ 
    &p_{q_p}=P_0-2\xi \psi^2 \cos\theta \,,
\end{align}
respectively. The expressions of $\mathcal{A}_n$ (with $n$ = 1,2,3,4), $\mathcal{B}_n$ (with $n$ = 1), $\mathcal{E}_n$ (with $n$ = 1), $\mathcal{F}_n$ (with $n$ = 1,2), $\mathcal{H}_n$ (with $n$ = 1,2,...,22), $\mathcal{I}_n$ (with $n$ = 1,2,...,18), $\mathcal{J}_n$ (with $n$ = 1,2,...,9), $\mathcal{K}_n (n=1,2,...,11)$, and $\mathcal{L}_n$ (with $n$ = 1,2,...,13) appearing in field variables for the elliptical Poiseuille flow (section 2, ESI$^{\dag}$) are simplified to $A_n$, $B_n$, $E_n$, $F_n$, $H_n$, $I_n$, $J_n$, $K_n$, and $L_n$, respectively, and are given in the ESI$^{\dag}$ (section 3).

The surface deformation of the particle until \textit{O}($\alpha^2$) is obtained as
\begin{align}
    f_{q_p}= \alpha \,f^{(1)}_{q_p}+\alpha^2 \,f^{(2)}_{q_p}
\end{align}
Here,
\begin{align}
  f^{(1)}_{q_p}=&\left[A_1 + E_1 - \frac{5}{4} A_3 \left( \frac{\cos2\theta}{2} - \cos2\phi \sin^2\theta \right)\right]\cos\theta -\frac{P_0}{2+3\Gamma} \,,\label{eq:f1 quadratic pp flow}\\
  f^{(2)}_{q_p}=& L_1 - L_2 \cos2\theta + L_3 \cos4\theta + L_4 \cos6\theta - (L_5 + L_6 \cos2\theta \nonumber\\&+ L_7 \cos4\theta) \cos2\phi \sin^2\theta + (L_8 + L_9 \cos2\theta) \cos4\phi \sin^4\theta \nonumber\\&+ P_0^2 L_{10} + P_0\left[L_{11} \cos 3\theta + \cos \theta ( L_{12} + L_{13} \cos 2\phi \sin^2 \theta )\right] \,.
\end{align}
The surface deformation is a function of the azimuthal angle ($\phi$) due to the asymmetry of the channel geometry about the $z$-axis. The point force until \textit{O}($\alpha^2$) in dimensional form is obtained as
\begin{align}
    \mathbf{F}_{q_p}=& 6\pi\mu V_0 R_{0}\left(\left[1+\frac{\psi^2}{3}\right]-\alpha P_0\left[\frac{1+\psi^2}{(2+3\Gamma)}\right]+\alpha^2 \left[\vphantom{\frac{1 + 2 \psi^2}{ (2 + 3 \Gamma)^2}}\mathcal{X}_1+\mathcal{
    X}_2 \psi^2\right.\right. \nonumber\\& \left.\left.+ \mathcal{X}_3 \psi^4+\mathcal{X}_4 \psi^6 +P_0^2\frac{1 + 2 \psi^2}{ (2 + 3 \Gamma)^2}\right]\right)\mathbf{\hat{k}} \,.\label{eq:force in pp flow}
\end{align}
\subsection{Simplification for the Hagen-Poiseuille flow}\label{subsec:Simplification for the Hagen Poiseuille flow}
Similar to the results obtained for the plane Poiseuille flow, all field quantities at different orders in $\alpha$ are obtained for the Hagen-Poiseuille flow by substituting $\psi_x=\psi_y=\psi$ in the corresponding expressions for the elliptical Poiseuille flow.
The components of undisturbed velocity and the pressure fields for the quadratic component of the Hagen-Poiseuille flow are obtained as
\begin{align}
    &{v_{q_h,r}}=-\xi^2\psi^2  \cos\theta \sin^2\theta\,,\\
    &v_{q_h,\theta}=\xi^2\psi^2 \sin^3\theta\,,\\
    &v_{q_h,\phi}=0\,\,\text{and }\\
    &p_{q_h}=P_0-4\xi\psi^2 \cos\theta \,,
\end{align}
The expressions of $\mathcal{A}_n$ (with $n$ = 1,2,3,4), $\mathcal{B}_n$ (with $n$ = 1), $\mathcal{E}_n$ (with $n$ = 1), $\mathcal{F}_n$ (with $n$ = 1,2), $\mathcal{H}_n$ (with $n$ = 1,2,...,22), $\mathcal{I}_n$ (with $n$ = 1,2,...,18), $\mathcal{J}_n$ (with $n$ = 1,2,...,9), $\mathcal{K}_n (n=1,2,...,11)$, and $\mathcal{L}_n$ (with $n$ = 1,2,...,13) are simplified to $\tilde{A}_n$, $\tilde{B}_n$, $\tilde{E}_n$, $\tilde{F}_n$, $\tilde{H}_n$, $\tilde{I}_n$, $\tilde{J}_n$, $\tilde{K}_n$, and $\tilde{L}_n$, respectively, and are given in the ESI$^{\dag}$ (section 4).

The surface deformation of the particle until \textit{O}($\alpha^2$) is obtained as
\begin{align}
    f_{q_h}= \alpha \,f^{(1)}_{q_h}+\alpha^2 \,f^{(2)}_{q_h}
\end{align}
Here,
\begin{align}
  f^{(1)}_{q_h}=&\left[\tilde{A}_1 + \tilde{E}_1 - \frac{5}{8} \tilde{A}_3  \cos2\theta\right]\cos\theta-\frac{P_0}{2+3\Gamma} \,,\\
  f^{(2)}_{q_h}=&\tilde{L}_1 - \tilde{L}_2 \cos2\theta + \tilde{L}_3 \cos4\theta + \tilde{L}_4 \cos6\theta+ P_0^2 \tilde{L}_{10} \nonumber\\&+ P_0\left[\tilde{L}_{11} \cos 3\theta + \cos \theta \tilde{L}_{12}\right]  \,.
\end{align}
The expression of surface deformation becomes simpler compared to the elliptical Poiseuille and plane Poiseuille flows, as it no longer depends on the azimuthal angle ($\phi$) due to the axisymmetric channel geometry about the $z$-axis. The point force until \textit{O}($\alpha^2$) in dimensional form is obtained as
\begin{align}
    \mathbf{F}_{q_h}=& 6\pi\mu V_0 R_{0}\left(\left[1+\frac{2\psi^2}{3}\right]-\alpha P_0\left[\frac{1+2\psi^2}{(2+3\Gamma)}\right]+\alpha^2 \left[\vphantom{\frac{1 + 4 \psi^2}{ (2 + 3 \Gamma)^2}}\mathcal{X}_1 + 2 \mathcal{X}_2 \psi^2 \right.\right. \nonumber\\& \left.\left.+  (2 \mathcal{X}_3 + \mathcal{X}_6)\psi^4 + 2 (\mathcal{X}_4 + \mathcal{X}_5) \psi^6+P_0^2\frac{1 + 4 \psi^2}{ (2 + 3 \Gamma)^2}\right]\right) \mathbf{\hat{k}} 
 \,. \label{eq:force in hp flow}
\end{align}
\subsection{Analyses of the velocity and deformed shape}\label{subsec:Analyses of the velocity and deformation}
In this section, we plot the velocity field and the deformed shape of the particle until \textit{O}($\alpha^2$) for all three Poiseuille flows, as shown in Fig. \ref{fig:schematic_pp_hp_flow}. The representation of streamplots is similar to that used by Chaithanya et al.,\citep{chaithanya2023active} who have analysed an active compound drop in a quadratic flow.
\subsubsection{Velocity field near the elastic particle}\label{subsubsec:Effect of elasticity on the velocity field}
\begin{figure*}
\centerline{\includegraphics[width=0.85\textwidth]{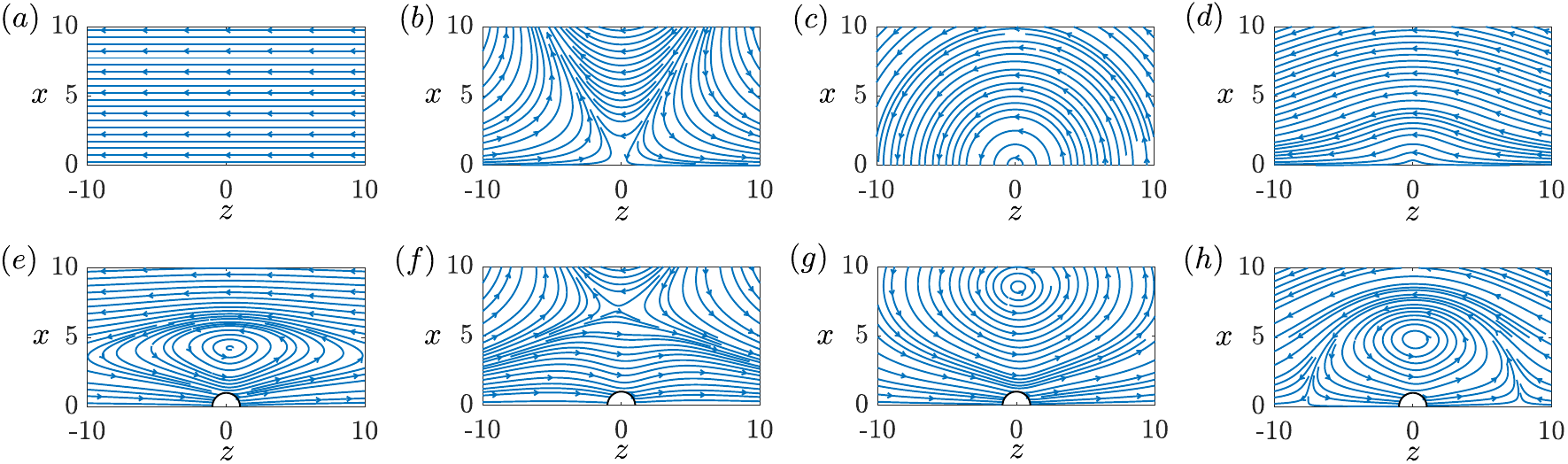}}
  \caption{Streamlines of (a) the quadratic component of elliptical Poiseuille flow and its irreducible components (b) $\boldsymbol{\gamma}_{q_e}$, (c) $\mathbf{Q}_{q_e}$ and (d) $\boldsymbol{\tau}_{q_e}$. (e)-(h) are the streamlines of the total velocity field (ambient plus disturbance field) corresponding to (a)-(d), at $\alpha=0.2$, $\Gamma=2$, $P_0=0.1$, $\psi_x=0.1$, and $\psi_y=0.08$.}
\label{fig:ambient + disturbance velocity field in lab frame_ep}
\end{figure*}

\begin{figure*}
 \centerline{\includegraphics[width=0.85\textwidth]{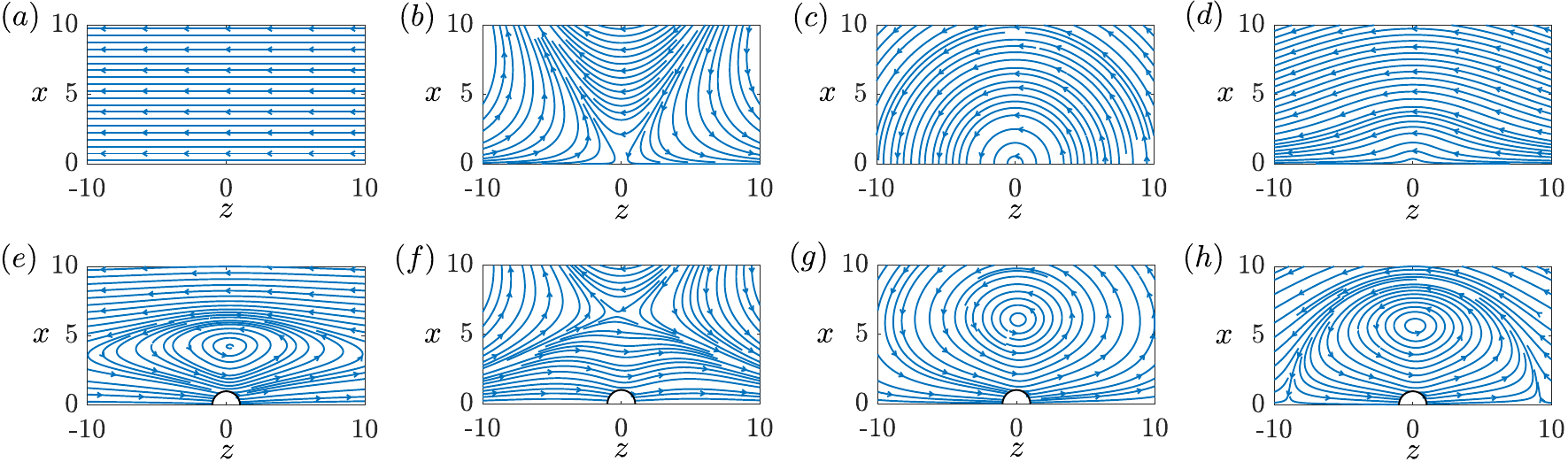}}
  \caption{Streamlines of (a) the quadratic component of plane Poiseuille flow, and its irreducible components (b) $\boldsymbol{\gamma}_{q_p}$, (c) $\mathbf{Q}_{q_p}$, and (d) $\boldsymbol{\tau}_{q_p}$. (e)-(h) are the streamlines of the total velocity field (ambient plus disturbance field) corresponding to (a)-(d), at $\alpha=0.2$, $\Gamma=2$, $P_0=0.1$, and $\psi=0.1$.}
\label{fig:ambient + disturbance velocity field in lab frame_pp}
\end{figure*}

\begin{figure*}
 \centerline{\includegraphics[width=0.7\textwidth]{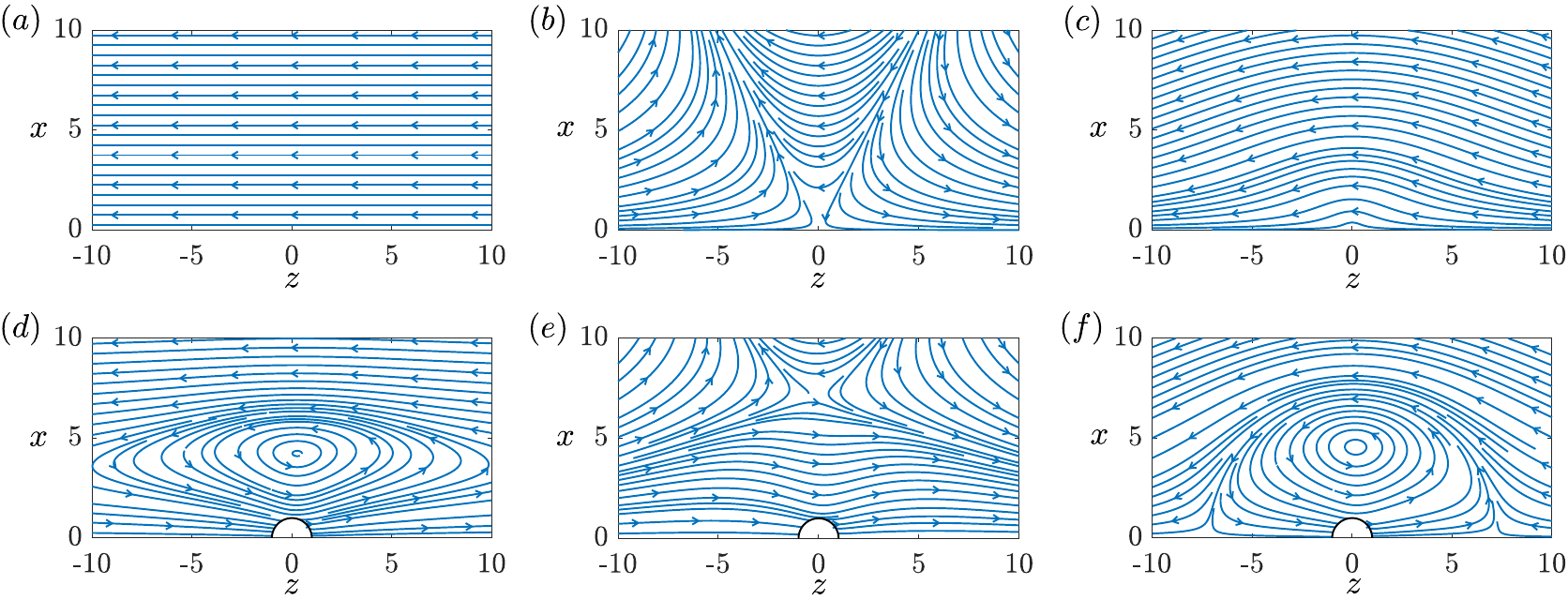}}
  \caption{Streamlines of (a) the quadratic component of Hagen-Poiseuille flow, and its irreducible components (b) $\boldsymbol{\gamma}_{q_h}$ and (c) $\boldsymbol{\tau}_{q_h}$. (d)-(f) are the streamlines of the total velocity field (ambient plus disturbance field) corresponding to (a)-(d), at $\alpha=0.2$, $\Gamma=2$, $P_0=0.1$, and $\psi=0.1$.}
\label{fig:ambient + disturbance velocity field in lab frame_hp}
\end{figure*}

We plot the streamlines in the laboratory frame near the particle in one half of the $xz$-plane for all three Poiseuille flows, since the flow is symmetric about the $z$-axis in the plane. The confinement ratio is considered the same on the $xz$-plane in all the three flows. The particle is constrained to move with $V_{max}$ in the quadratic flow, as well as in its irreducible components. The streamlines of the quadratic flow corresponding to the elliptical Poiseuille flow ($\mathbf{v}_{q_e}$) are shown in Fig. \ref{fig:ambient + disturbance velocity field in lab frame_ep}(a) and those of its irreducible components $\boldsymbol{\gamma}_{q_e}$, $\mathbf{Q}_{q_e}$ and $\boldsymbol{\tau}_{q_e}$ are shown in Figs. \ref{fig:ambient + disturbance velocity field in lab frame_ep}(b), (c), and (d), respectively.
The streamlines of the total velocity field (ambient plus disturbance field) corresponding to $\mathbf{v}_{q_e}$, and those corresponding to $\boldsymbol{\gamma}_{q_e}$, $\mathbf{Q}_{q_e}$ and $\boldsymbol{\tau}_{q_e}$ are shown in Figs. \ref{fig:ambient + disturbance velocity field in lab frame_ep}(e)-(h). The streamlines of the undisturbed velocity field corresponding to $\boldsymbol{\gamma}_{q_e}$ [Fig. \ref{fig:ambient + disturbance velocity field in lab frame_ep}(b)] exhibit an extensile-compressive hexapolar structure. The structure exhibits six distinct principal directions on the $xz$-plane, with the extensional and compressional states aligned along these principal axes in an alternating manner. The streamlines of the total velocity field (ambient plus disturbance field) associated with $\boldsymbol{\gamma}_{q_e}$ are presented in Fig. \ref{fig:ambient + disturbance velocity field in lab frame_ep}(f). For $\mathbf{Q}_{q_e}$, the streamlines of the undisturbed velocity field [Fig. \ref{fig:ambient + disturbance velocity field in lab frame_ep}(c), upper half shown] are circular, featuring opposite axes of rotation in the upper and lower halves of the $xz$-plane. The streamlines of the corresponding total velocity field are plotted in Fig. \ref{fig:ambient + disturbance velocity field in lab frame_ep}(g). For $\boldsymbol{\tau}_{q_e}$, the streamlines of the undisturbed velocity field [Fig. \ref{fig:ambient + disturbance velocity field in lab frame_ep}(d)] resembles uniform flow past a rigid sphere, with the associated total velocity field shown in Fig. \ref{fig:ambient + disturbance velocity field in lab frame_ep}(h). Similarly, we plot the streamlines of the quadratic flow, its irreducible components (top row in the concerned figure), and those of the total velocity field corresponding to the quadratic flow and its irreducible components (bottom row in the concerned figure) for the plane Poiseuille and Hagen–Poiseuille flows in Figs. \ref{fig:ambient + disturbance velocity field in lab frame_pp} and \ref{fig:ambient + disturbance velocity field in lab frame_hp}, respectively. 
The streamlines of the total velocity field corresponding to quadratic flow [Figs. \ref{fig:ambient + disturbance velocity field in lab frame_ep}(e), \ref{fig:ambient + disturbance velocity field in lab frame_pp}(e), and \ref{fig:ambient + disturbance velocity field in lab frame_hp}(d)] are almost similar in all three flows. A similar observation also holds for the streamlines corresponding to $\boldsymbol{\gamma}$ [Figs. \ref{fig:ambient + disturbance velocity field in lab frame_ep}(f), \ref{fig:ambient + disturbance velocity field in lab frame_pp}(f), and \ref{fig:ambient + disturbance velocity field in lab frame_hp}(e)]. The component $\mathbf{Q}$ is zero in the Hagen-Poiseuille flow, while it is non-zero in the elliptical Poiseuille [Fig. \ref{fig:ambient + disturbance velocity field in lab frame_ep}(c)] and plane Poiseuille [Fig. \ref{fig:ambient + disturbance velocity field in lab frame_pp}(c)] flows. The centre of the circulation zone in the streamlines of total velocity field, corresponding to $\mathbf{Q}$, is located away from the particle in the elliptical Poiseuille flow compared to that in the plane Poiseuille flow, as shown in Figs. \ref{fig:ambient + disturbance velocity field in lab frame_ep}(g) and \ref{fig:ambient + disturbance velocity field in lab frame_pp}(g), respectively. Further, the location of the centre of the circulation zone in the total velocity field, corresponding to $\boldsymbol{\tau}$, is similar for the elliptical Poiseuille [Fig. \ref{fig:ambient + disturbance velocity field in lab frame_ep}(h)], plane Poiseuille flow [Fig. \ref{fig:ambient + disturbance velocity field in lab frame_pp}(h)], and Hagen-Poiseuille [Fig. \ref{fig:ambient + disturbance velocity field in lab frame_hp}(f)] flows.

\subsubsection{Deformed shape of the elastic particle}\label{subsubsec:Effect of elasticity on the deformed shape}
\begin{figure*}
 \centerline{\includegraphics[width=0.8\textwidth]{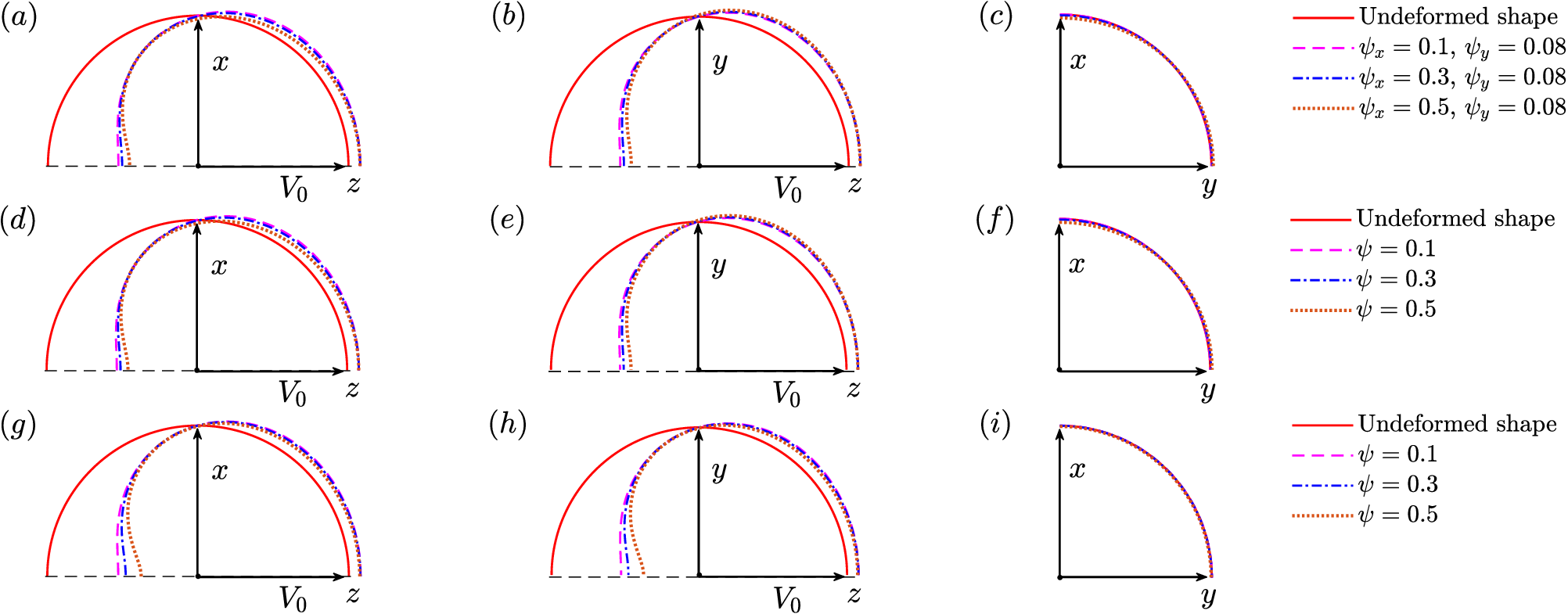}}
  \caption{Deformed shape of the sphere for different confinement ratios, (a)-(c) for elliptical Poiseuille flow, (d)-(f) for plane Poiseuille flow and (g)-(i) for Hagen-Poiseuille flow plotted on the $xz$-, $yz$- and $xy$-planes, respectively, at $\alpha=0.2$, $P_0=0.1$, and $\Gamma=2$.}
\label{fig:deformation psi variation_ep pp hp}
\end{figure*}

\begin{figure*}
 \centerline{\includegraphics[width=0.82\textwidth]{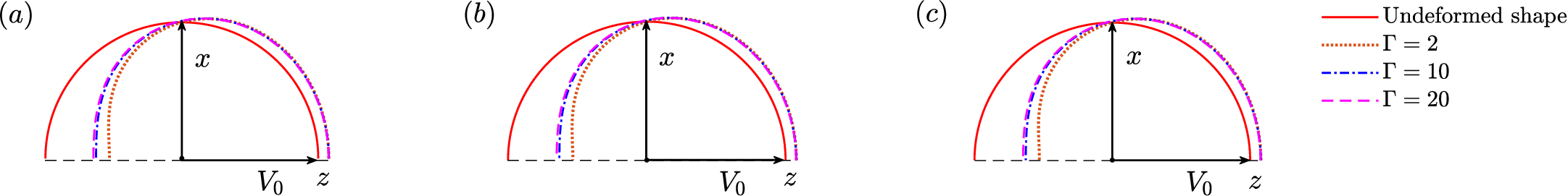}}
  \caption{Deformed shape of the sphere for different $\Gamma$, (a) for elliptical Poiseuille flow at $\psi_x=0.1$ and $\psi_y=0.08$, (b) for plane Poiseuille flow at $\psi_x=\psi=0.1$ and $\psi_y=0$ and (c) for Hagen-Poiseuille flow at $\psi_x=\psi_y=\psi=0.1$, plotted on the $xz$-plane at $\alpha=0.2$, and $P_0=0.1$. }
\label{fig:deformation gamma variation_ep pp hp}
\end{figure*}

In this section, we discuss the deformed shape of the particle. The shape is plotted in one half of the $xz$-, $yz$- and $xy$-planes for the elliptical Poiseuille [Figs. \ref{fig:deformation psi variation_ep pp hp}(a)-(c)], plane Poiseuille [Figs. \ref{fig:deformation psi variation_ep pp hp}(d)-(f)] and Hagen-Poiseuille flows [Figs. \ref{fig:deformation psi variation_ep pp hp}(g)-(i)], for different confinement ratios.
As expected, a larger confinement ratio results in more deformation of the particle compared to that for a lower confinement ratio. In the elliptical Poiseuille and plane Poiseuille flows, the deformed shape is not axisymmetric about the $z$-axis as seen in Figs. \ref{fig:deformation psi variation_ep pp hp}(a)-(c) and (d)-(f), respectively. Lack of axisymmetry is due to the asymmetry of the surrounding flow field. The shape is axisymmetric about the $z$-axis in the Hagen-Poiseuille flow [Figs. \ref{fig:deformation psi variation_ep pp hp}(g)-(i)]. The dip observed on the left side of the particle on the $xz$- (or $yz$-) plane in all three Poiseuille flows results from the point force applied at the centre of the particle. The deformed shape of the particle also depends on the material property ($\Gamma$, defined following eqn (\ref{eq:nondimensional variables})). The shape is plotted on the $xz$-plane for different $\Gamma$ for the elliptical Poiseuille [Fig. \ref{fig:deformation gamma variation_ep pp hp}(a)], plane Poiseuille [Fig. \ref{fig:deformation gamma variation_ep pp hp}(b)], and Hagen-Poiseuille [Fig. \ref{fig:deformation gamma variation_ep pp hp}(c)] flows.
As is evident from the Fig. \ref{fig:deformation gamma variation_ep pp hp}, the particle exhibits reduced volumetric deformation for larger values of $\Gamma$, compared to lower values for all three Poiseuille flows.

\subsubsection{Deformed shape of the elastic particle and a drop in a plane Poiseuille flow}\label{subsec:comparison with drop}

\begin{figure}[h]
\centerline{\includegraphics[width=0.55\textwidth]{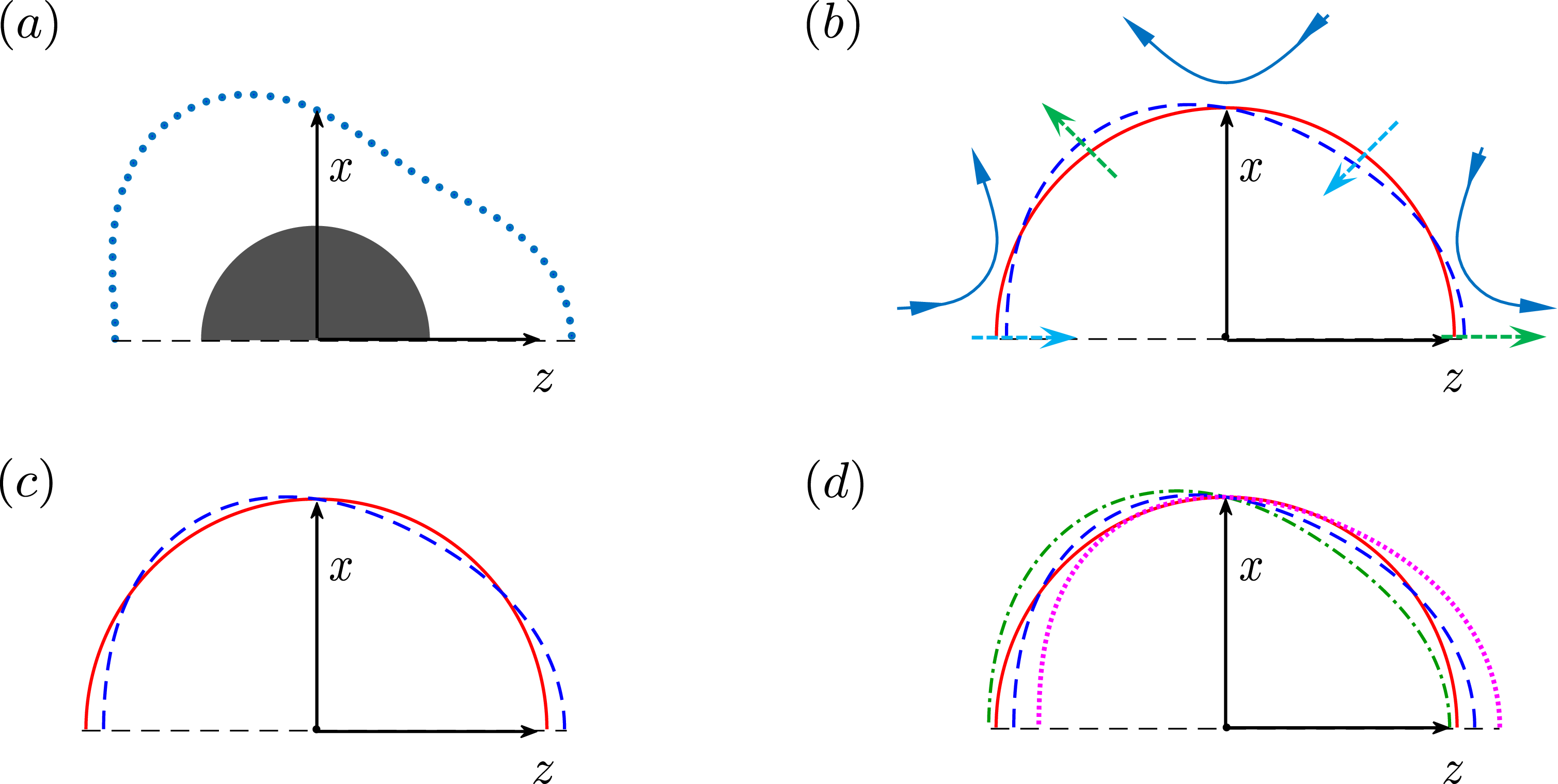}}
  \caption{(a) Deformed shape of the compound drop at zero activity.\citep{chaithanya2023active} (b)  Deformed shape of the elastic particle as it translates with $V_{max}$ in plane Poiseuille flow, corresponding to (b) $\boldsymbol{\gamma}$, and (c) $\boldsymbol{\gamma}$ and $\boldsymbol{\tau}$. (d) Variation in the deformed shape of the particle for different $V_{0}$ relative to $V_{max}$; dash-dot for $V_0= 0.8 V_{max}$, dashed for $V_0= V_{max}$, and dotted for $V_0= 1.2 V_{max}$. Parameters: $\alpha=0.4$, $\psi=0.5$, $\Gamma=2$, and $P_0=0$. In (b), the arrows (blue) outside the deformed shape represent $\boldsymbol{\gamma}$, and arrows with dotted lines represent the compression (light blue) or extension (green) of the particle depending on the direction of the arrows. The undeformed shape is shown using a red solid line.}
\label{fig:comparison with drop}
\end{figure}

In the literature, it has been reported that an incompressible simple/active compound spherical drop in a plane Poiseuille flow at the centreline, deforms to a three-lobe shape at \textit{O}($Ca$), \citep{nadim1991motion,chaithanya2023active} due to the hexapolar component of the flow. Here, $Ca$ is the capillary number defined as the ratio of viscous forces to surface tension forces. The three-lobe shape at \textit{O}($Ca$) of the active compound drop at zero activity is shown in Fig. \ref{fig:comparison with drop}(a).\citep{chaithanya2023active} As the activity strength increases, the shape changes from the three-lobe structure. We analyse and compare the leading order shape dynamics of the active compound drop with that of the elastic particle translating in the same flow, with the reference pressure set to zero. At the centreline, the Poiseuille flow reduces to a linear combination of uniform and quadratic flow components. The results presented up to section \ref{subsubsec:Effect of elasticity on the deformed shape} correspond to the particle translating in only the quadratic component of the flow. To the leading order, we can superimpose the deformation of the elastic particle in the quadratic component with that in the uniform component to get the particle's shape for the comparison. The superposition is owing to the linearity of the boundary conditions; \textit{O}($1$) velocity boundary conditions [eqn (\ref{eq:vel_v0_bc1}) and (\ref{eq:vel_v0_bc2})] and \textit{O}($\alpha$) stress boundary conditions [eqn (\ref{eq:Oalpha_stressbc_r})-(\ref{eq:Oalpha_stressbc_phi})].
We carried out the calculations for the particle translating in the uniform and quadratic components, with the rescaled velocity as $V_{max}$. The expressions for the surface deformation in both the uniform and quadratic components are given in Appendix D.
The elastic particle also deforms to the three-lobe shape, in the flow field corresponding to $\boldsymbol{\gamma}$ [Fig. \ref{fig:comparison with drop}(b)] and a nearly three-lobe shape in the complete flow (with additional deformation arising from $\boldsymbol{\tau}$) [Fig. \ref{fig:comparison with drop}(c)]. Further, the particle does not deform in the flow field arising from $\mathbf{Q}$, same as in the drop at \textit{O}($Ca$).\citep{nadim1991motion,chaithanya2023active} 
Interestingly, the deformed shape of the elastic particle is also influenced by the velocity of the particle itself. The three-lobe shape transforms into a distinct shape when the particle translates at a velocity different from the $V_{max}$, as shown in Fig. \ref{fig:comparison with drop}(d). Such a shape change from the three-lobe shape is also reported for the active compound drop with an increase in activity strength.\citep{chaithanya2023active}  Therefore, the manipulation of the elastic particle embedded with a magnetic bead using an external magnetic field is analogous to the manipulation of the active compound drop using the strength of the activity. We also note that the reference pressure does not affect the deformed shape of the drop owing to its incompressibility, whereas for the elastic particle considered here, a nonzero reference pressure induces compression in the particle without modifying its leading-order deformed shape. 

\section{Summary}\label{sec:summary}
We build on the theoretical framework developed in Verma et al.,\cite{verma2025dynamics} where the dynamics of an internally actuated elastic particle has been analysed as it translates parallel to a rigid wall in a quiescent fluid. The particle models the response of a spherical polymer bead embedded with a single magnetic particle at its undeformed centre. The external magnetic force/torque acting on such a bead can be effectively modelled as a localised force/torque, using a point force/torque approach, given that the embedded magnetic particle is much smaller in size compared to the bead size. In the present work, the elastic particle is constrained to translate in a general unbounded quadratic flow using a point force/torque approach. We consider that the particle is weakly elastic and compressible in nature. The dynamics and morphology of the particle are analysed in the Stokes limit, neglecting gravitational effects. The fluid and the particle are modelled using the Stokes equations and the Navier elasticity equations, respectively. The results for the particle motion in the general quadratic flow are presented in section \ref{sec:results} and the results are simplified further for the particle translating with $V_{max}$ along the channel centreline in the quadratic component of: 1) elliptical Poiseuille flow discussed in section \ref{subsec:Simplification for the elliptical Poiseuille flow}, 2) plane Poiseuille flow discussed in section \ref{subsec:Simplification for the plane Poiseuille flow} and 3) Hagen-Poiseuille flow discussed in section \ref{subsec:Simplification for the Hagen Poiseuille flow}. The velocity, pressure, displacement, point force, and point torque are obtained until \textit{O}($\alpha^2$). In the general quadratic flow, the effects of elasticity appear as an external force at \textit{O}($\alpha$) and is aligned with the velocity of the particle ($z$-axis) and the force at \textit{O}($\alpha^2$) has components along all three axes given in eqn (\ref{eq:force in general quadratic}). In contrast, the external force until \textit{O}($\alpha^2$) is aligned with the particle velocity in all three Poiseuille flows given in eqn (\ref{eq:force in ep flow}), (\ref{eq:force in pp flow}) and (\ref{eq:force in hp flow}). In the general quadratic flow, the external torque due to the elastic effects appears at \textit{O}($\alpha$) and \textit{O}($\alpha^2$) given in eqn (\ref{eq:torque in general quadratic}), while the torque is zero in all three Poiseuille flows. The streamlines for the quadratic flow and its irreducible components, and those of the total velocity field (ambient plus disturbance fields) are plotted in a laboratory frame for the elliptical Poiseuille [Figs. \ref{fig:ambient + disturbance velocity field in lab frame_ep}(a)-(h)], plane Poiseuille [Figs. \ref{fig:ambient + disturbance velocity field in lab frame_pp}(a)-(h)] and Hagen-Poiseuille [Figs. \ref{fig:ambient + disturbance velocity field in lab frame_hp}(a)-(f)] flows. The $\boldsymbol{\gamma}$ component of the Poiseuille flows exhibits the extensile-compressive hexapolar structure, which is responsible for generating the three-lobe shape of the deformable particle when subjected to this flow component. The $\mathbf{Q}$ component leads to a deformation of the elastic particle due to the compressibility of the particle.
The $\boldsymbol{\tau}$ component leads to a deformation of the particle similar to that experienced by the particle in a uniform flow. The final deformed shape of the particle is determined by the force distribution arising from the equilibrium of hydrodynamic forces, the point force, and the point torque. The variation in the shape of the particle is plotted for different confinement ratios ($\psi$) and material property ($\Gamma$), as shown in Figs. \ref{fig:deformation psi variation_ep pp hp} and \ref{fig:deformation gamma variation_ep pp hp}, respectively. The deformation exhibited by the particle increases with the increase in $\psi$ or decrease in $\Gamma$ and vice versa.\\
We also compare the leading-order deformed shape of the elastic particle at the centreline in a plane Poiseuille flow with that of a drop in the same flow.\citep{chaithanya2023active} While the drop exhibits a three-lobe shape in the flow, the elastic particle can also exhibit a similar shape depending on the velocity ratio $V_0/V_{max}$, as shown in [Fig. \ref{fig:comparison with drop}(d)]. The shape variation of the elastic particle with the velocity ratio is analogous to that of an active compound drop with changing activity strength. The deformed shape will also depend on the force distribution acting on the particle. In a Hagen-Poiseuille flow, an incompressible neo-Hookean sphere subjected to an axial body force and moving along the centerline deforms into a bullet-like shape, an anti-bullet shape, or retains its spherical shape, depending on the magnitude of the external body force and viscous forces.\citep{finney2024impact} Such a shape will not be observed in the elastic particle considered here because of a change in the force distribution arising from the balance of the point force/torque and the hydrodynamic forces. A thorough understanding of the force distribution acting on a deformable particle is essential for studying the particle dynamics. Our analysis plays a crucial role in understanding the fluid-particle interaction and in determining the steady-state morphology of an elastic particle translating in quadratic flows.  The analytical framework developed in this study can be extended to explore the behaviour of viscoelastic/nonlinear elastic polymer beads in quadratic flows. 
%\begin{acknowledgments}
%Acknowledge 
%\end{acknowledgments}
\section{Data availability}
$^{\dag}$Electronic supplementary information (ESI): The velocity, pressure fields at \textit{O}($1$) and displacement field at \textit{O}($\alpha$) for the general quadratic flow; simplified results for the quadratic components of the Poiseuille flows.

$^{\ddag}$Electronic supplementary information (ESI): The constants needed for the determination of the disturbance velocity and pressure fields at \textit{O}($\alpha$); the displacement field at \textit{O}($\alpha^2$); the disturbance velocity and pressure fields at \textit{O}($\alpha^2$) are available in a Wolfram Mathematica sheet.
\appendix
\section{Series solutions to the Stokes and continuity equations}\label{appsec:Series solution to Stokes continuity equation}
The radial ($v_r$), tangential ($v_\theta$), and azimuthal ($v_\phi$) components of the velocity field are given by
 \begin{align}
v_r=&\sum_{n=1}^{\infty} \sum_{m=0}^{n} \left[\left (\frac{\xi^{-n}(n+1)}{2(2n-1)}P_n^m[a[m,n] \cos m\phi + \tilde{a}[m,n] \sin m\phi]\right)\right.\nonumber\\& \left.- 
\vphantom{\frac{(n+1)}{2(2n-1)}}\left(\xi^{-n-2}(n+1)P_n^m[b[m,n] \cos m\phi +\tilde{b}[m,n] \sin m\phi]\right)\right] \,,\label{eq:vr series}
\end{align}
\begin{align}
v_{\theta}=&\sum_{n=1}^{\infty} \sum_{m=0}^{n}\left[\left(\frac{\xi^{-n}(n-2)}{2n(2n-1)(2n+1) \sin \theta}[(n+1)(n+m)P_{n-1}^m\right.\right.\nonumber\\&\left.\left.-n(n-m+1)P_{n+1}^m]\vphantom{\frac{\xi^{-n}(n-2)}{2n(2n-1)(2n+1) \sin \theta}}\right)[a[m,n] \cos m\phi + \tilde{a}[m,n] \sin m\phi]\right.\nonumber\\&\left.-\left (\frac{\xi^{-n-2}}{(2n+1) \sin \theta} [(n+1)(n+m)P_{n-1}^m-n(n-m+1)P_{n+1}^m]\right)\right.\nonumber\\&\left. \times [b[m,n] \cos m\phi + \tilde{b}[m,n] \sin m\phi]\right.\nonumber\\&\left.+
\left (\frac{\xi^{-n-1} m}{\sin \theta}P_n^m[-v[m,n] \sin m\phi +\tilde{v}[m,n] \cos m\phi]\vphantom{\frac{\xi^{-n-1} m}{\sin \theta}}\right)\right]\label{eq:vtheta series}
\end{align}
and
\begin{align}
v_{\phi}=&\sum_{n=1}^{\infty} \sum_{m=0}^{n}\left[\left(-\frac{\xi^{-n}(n-2)m}{2n(2n-1)\sin \theta} P_n^m [-a[m,n] \sin m\phi \right.\right.\nonumber\\&\left.\left.+ \tilde{a}[m,n] \cos m\phi]\vphantom{\frac{\xi^{-n}(n-2)m}{2n(2n-1)\sin \theta}}\right)\right.\nonumber\\&\left.+
\left (\frac{\xi^{-n-2} m}{\sin \theta} P_n^m [-b[m,n] \sin m\phi +\tilde{b}[m,n] \cos m\phi]\right)\right.\nonumber\\&\left.+
\left (\frac{\xi^{-n-1}}{(2n+1)\sin \theta}[(n+1)(n+m)P_{n-1}^m -n(n-m+1)P_{n+1}^m]\right)\right.\nonumber\\&\left.\times (v[m,n] \cos m\phi +\tilde{v}[m,n] \sin m\phi)\vphantom{\frac{\xi^{-n}(n-2)m}{2n(2n-1)\sin \theta}}\right]\,,\label{eq:vphi series}
\end{align}
respectively. The pressure field is given by
\begin{align}
p&= \sum_{n=1}^{\infty} \sum_{m=0}^{n}\left[\xi^{-n-1}P_n^m(a[m,n] \cos m\phi + \tilde{a}[m,n] \sin m\phi)\right] \,. \label{eq:press series}
\end{align} 
Above, $P^m_n$ denotes $P^m_n(\cos \theta)$, the associated Legendre polynomial of order $m$ and degree $n$. The constants, $a[m,n]$, $\tilde{a}[m,n]$, $b[m,n]$, $\tilde{b}[m,n]$, $v[m,n]$, and $\tilde{v}[m,n]$ are determined from the no-slip and no-penetration boundary conditions at the surface of the particle.
\section{Series solutions to the Navier elasticity equations}\label{appsec:Series solution to Navier elasticity equation}
The radial ($u_r$), tangential ($u_{\theta}$), and azimuthal ($u_\phi$) components of the displacement field are given by
\begin{align}
u_r=&\frac{1}{4 \pi \xi}\left[\vphantom{\frac{1}{4 \pi \xi}}F_{z} \cos \theta + F_{x} \cos \phi \sin \theta + F_{y} \sin \theta \sin \phi\right]\nonumber \\&+\sum_{n=0}^{\infty} \sum_{m=0}^{n}\left[\xi^{n+1}P_n^m(b1[m,n] \cos m\phi +b0[m,n] \sin m\phi)\right] \nonumber \\&+\sum_{n=1}^{\infty} \sum_{m=0}^{n}\left[\xi^{n-1}P_n^m(c1[m,n] \cos m\phi +c0[m,n] \sin m\phi)\right]\,, \label{eq:ur series}
\end{align}
\begin{align}
u_{\theta}=&\frac{(3 + \Gamma)}{8 \pi (2 + \Gamma) \xi}\left(\vphantom{\frac{(3 + \Gamma)}{8 \pi (2 + \Gamma) \xi}} -F_{z} \sin \theta + \cos \theta [F_{x} \cos \phi + F_{y} \sin \phi ]\right)\nonumber \\&+\frac{1}{8 \pi \xi^2}\left(\vphantom{\frac{1}{8 \pi \xi^2}}T_{y} \cos \phi - T_{x} \sin \phi \right)+\frac{1}{\sin \theta}\left[\sum_{n=1}^{\infty} \sum_{m=0}^{n}\left(\frac{\xi^n m(2n+1)}{n (n+1)}\right.\right.\nonumber \\&\left.\left. \times P_n^m [-a1[m,n] \sin m\phi +a0[m,n] \cos m\phi]\vphantom{\frac{\xi^n m(2n+1)}{n (n+1)}}\right)\vphantom{\sum_{n=1}^{\infty}}\right] \nonumber \\&+ 
\frac{1}{\sin \theta}\left[\sum_{n=1}^{\infty} \sum_{m=0}^{n}\left(\frac{\xi^{n+1} [\Gamma(n+3)+(n+5)]}{(2n+1)(n+1)[\Gamma\, n +(n-2)]} \right.\right.\nonumber \\&\left.\left. \times [n(n-m+1) P_{n+1}^m-(n+1)(n+m)P_{n-1}^m]\vphantom{\frac{\xi^{n+1} [\Gamma(n+3)+(n+5)]}{(2n+1)(n+1)[\Gamma\, n +(n-2)]}}\right)\right.\nonumber \\&\left. \times(b1[m,n] \cos m\phi +b0[m,n] \sin m\phi)\vphantom{\sum_{n=1}^{\infty}}\right] \nonumber\\&+\frac{1}{\sin \theta}\left[\sum_{n=1}^{\infty} \sum_{m=0}^{n}\left( \frac{\xi^{n-1}}{(2n+1)n}[n(n-m+1) P_{n+1}^m-(n+1)\right.\right.\nonumber \\&\left.\left. \times (n+m)P_{n-1}^m]\vphantom{\frac{\xi^{n-1}}{(2n+1)n}}\right)(c1[m,n] \cos m\phi +c0[m,n] \sin m\phi)\vphantom{\sum_{n=1}^{\infty}}\right]
\end{align}
and
\begin{align}
u_{\phi}=&\frac{(3 + \Gamma)}{8 \pi (2 + \Gamma) \xi} \left(\vphantom{\frac{(3 + \Gamma)}{8 \pi (2 + \Gamma) \xi}}F_{y} \cos \phi - F_{x} \sin \phi \right)\nonumber\\&+\frac{1}{8 \pi \xi^2} \left(\vphantom{\frac{1}{8 \pi \xi^2}}T_{z} \sin \theta - \cos \theta [T_{x} \cos \phi + T_{y} \sin \phi ]\right)\nonumber\\&+\frac{1}{\sin \theta}\left[\sum_{n=0}^{\infty} \sum_{m=0}^{n}\left(\frac{-\xi^n}{(n+1)} (n-m+1) P_{n+1}^m \right.\right.\nonumber\\&\left.\left. \times [a1[m,n] \cos m\phi +a0[m,n] \sin m\phi]\vphantom{\frac{-\xi^n}{(n+1)}}\right) + \sum_{n=1}^{\infty} \sum_{m=0}^{n}\left(\frac{\xi^n}{n}(n+m)\right.\right.\nonumber\\&\left.\left. \times P_{n-1}^m[a1[m,n] \cos m\phi +a0[m,n] \sin m\phi]\vphantom{\frac{\xi^n}{n}} \right)\vphantom{\sum_{n=1}^{\infty}}\right] \nonumber\\&+\frac{1}{\sin \theta}\left[\sum_{n=0}^{\infty} \sum_{m=0}^{n}\left( \frac{\xi^{n+1} m [\Gamma(n+3)+(n+5)]}{(n+1)[\Gamma \, n +(n-2)]} P_{n}^m \right.\right.\nonumber\\&\left.\left. \times[-b1[m,n] \sin m\phi +b0[m,n] \cos m\phi]\vphantom{\frac{\xi^{n+1} m [\Gamma(n+3)+(n+5)]}{(n+1)[\Gamma \, n +(n-2)]}}\right)\vphantom{\sum_{n=0}^{\infty}}\right]+\frac{1}{\sin \theta}\left[\sum_{n=1}^{\infty} \sum_{m=0}^{n} \right.\nonumber\\&\left.\left (\frac{\xi^{n-1} m}{n} P_{n}^m [-c1[m,n] \sin m\phi +c0[m,n] \cos m\phi]\right )\vphantom{\sum_{n=1}^{\infty}}\right],\label{eq:uphi series}
\end{align}
respectively. Here, the constants, $a1[m,n]$, $a0[m,n]$, $b1[m,n]$, $b0[m,n]$, $c1[m,n]$, and $c0[m,n]$ are determined from the stress continuity at the particle surface.

\section{Formulation of point force/torque in the presence of body force}\label{appsec:Formulation of point force/torque in the presence of body force}

\begin{figure}[h]
\centerline{\includegraphics[width=0.55\textwidth]{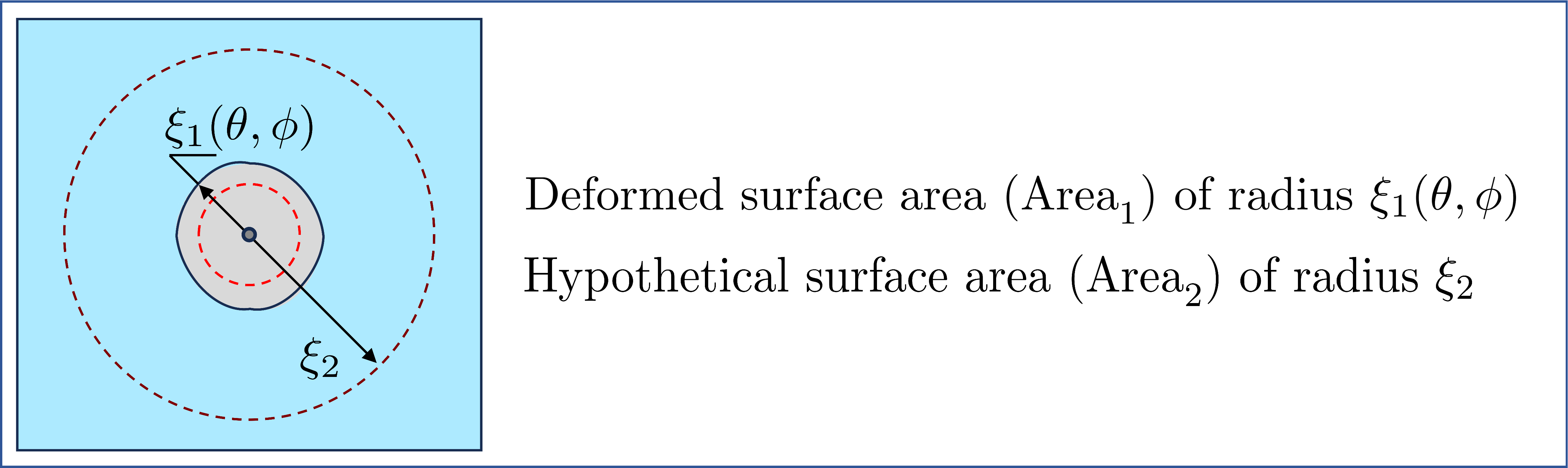}}
  \caption{Schematic of the deformed sphere, showing different regions of interest. Volume of deformed sphere ($\text{Vol}_1$) of radius $\xi_1(\theta, \phi)$ and outside the deformed sphere ($\text{Vol}_2$) between radial distance $\xi_1(\theta, \phi)$ and $\xi_2$.}
\label{appfig:schematic_point_force_fluid_solid_stress}
\end{figure}

Consider that the nondimensional body force term, $\mathbf{K}_p=\rho_p \mathbf{g} R_0/G$, where $\rho_p$ and $\mathbf{g}$ are the particle density and gravity, respectively, is also included in the Navier elasticity equations. The volume integral of the equation yields
\begin{align}
    \int_{\text{Vol}_1} (\mathbf{\nabla} \cdot  \boldsymbol{\hat{\sigma}})\,dV=-\left(\mathbf{F}+\int_{\text{Vol}_1} \mathbf{K}_p \,dV\right)\,. \label{appeq:volume integral of elasticity equation having body force term}
\end{align}
Here, $\text{Vol}_1$ is the volume of the deformed sphere of area $\text{Area}_1$, as shown in Fig. \ref{appfig:schematic_point_force_fluid_solid_stress}. 
The term on the left-hand side of eqn (\ref{appeq:volume integral of elasticity equation having body force term})can be converted to an area integral using Gauss's divergence theorem. The resulting equation after applying the stress continuity at the deformed interface is obtained as
\begin{align}
    \alpha \int_{\text{Area}_1} ( \boldsymbol{{\sigma}} \cdot \mathbf{\hat{n}})\,dA= -\left(\mathbf{F}+\int_{\text{Vol}_1} \mathbf{K}_p \, \xi^2 \sin \theta\,d\xi\, d\theta\, d\phi\right)\,. \label{appeq:fluid stress integral over deformed surface}
\end{align}
Here, $\text{Area}_1$ itself is a part of the solution. The fluid stress distribution over $\text{Area}_1$ can be expressed in terms of the body force within volume $\text{Vol}_2$ enclosed between $\text{Area}_1$ and a hypothetical surface area ($\text{Area}_2$) outside the deformed surface. Accordingly, eqn (\ref{appeq:fluid stress integral over deformed surface}) can be written as
\begin{align}
    \alpha \int_{\text{Area}_2} ( \boldsymbol{{\sigma}} \cdot \mathbf{\hat{n}}_2)\,dA=& -\left(\mathbf{F}+\int_{\text{Vol}_1} \mathbf{K}_p \, \xi^2 \sin \theta\,d\xi\, d\theta\, d\phi\,+ \right.\nonumber\\&\left.\alpha \int_{\text{Vol}_2} \mathbf{K}_f \, \xi^2 \sin \theta\,d\xi\, d\theta\, d\phi \right)\,. \label{appeq:final intergral expression of calculating force form the fluid and solid stress}
\end{align}
Here, $\mathbf{\hat{n}}_2$ is the unit outward normal vector to the $\text{Area}_2$ and $\mathbf{K}_{f}=\rho_f {R_0^2}\mathbf{g}/(\mu V_0)$. To simplify the calculation, $\text{Area}_2$ is considered as the surface area of a sphere. Note that the point force at the leading-order induces the stress inside the particle at \textit{O}($\alpha$), and the point force at \textit{O}($\alpha$) induces stress at \textit{O}($\alpha^2$). As is evident from eqn (\ref{appeq:final intergral expression of calculating force form the fluid and solid stress}), in the absence of body force ($\mathbf{K}_p=\mathbf{K}_f=0$), as in the presented work, the integral of the $\boldsymbol{{\sigma}} \cdot \mathbf{\hat{n}}_2$ over $\text{Area}_2$ give the external point force directly.
Similarly, one can also obtain the expression for point torque. The resulting expression is obtained as
\begin{align}
    \alpha \int_{\text{Area}_2} (\mathbf{x} \times [\boldsymbol{{\sigma}} \cdot \mathbf{\hat{n}}_2])\,dA=& -\left(\mathbf{T}+\int_{\text{Vol}_1} [\mathbf{x} \times \mathbf{K}_p] \, \xi^2 \sin \theta\,d\xi\, d\theta\, d\phi\,+ \right.\nonumber\\&\left.\alpha \int_{\text{Vol}_2} [\mathbf{x} \times \mathbf{K}_f] \, \xi^2 \sin \theta\,d\xi\, d\theta\, d\phi \right)\,. \label{appeq:final intergral expression of calculating torque form the fluid and solid stress}
\end{align}
Here, $\mathbf{x}$ is the non-dimensional position vector measured from the centre of the undeformed particle and $\xi=|\mathbf{x}|$. In the absence of body force ($\mathbf{K}_p=\mathbf{K}_f=0$), the external point torque can be determined directly from the fluid stress distribution using eqn (\ref{appeq:final intergral expression of calculating torque form the fluid and solid stress}). 

\section{Surface deformation ($f^{(1)}$) in uniform and quadratic component of plane Poiseuille flow}\label{appsec:Surface deformation in uniform and quadratic component of plane Poiseuille flow}
The velocity of the particle ($V_{0}\mathbf{\hat{k}}$) along the centreline in the plane Poiseuille flow is decomposed into the velocity in the uniform flow component, $V_{0,un}\mathbf{\hat{k}}$, and the velocity in the quadratic flow component, $V_{0,q}\mathbf{\hat{k}}$, such that $V_{0}=V_{0,un}+V_{0,q}$. 
The flow velocity inside the channel is maximum at the centreline ($V_{max}$), and is chosen as the velocity scale for the analysis below. The surface deformation of the particle at the leading-order in a uniform component of magnitude $V_{max}$ is represented as $f^{(1)}_{uni,p}$ and is obtained as
\begin{align}
    f^{(1)}_{uni,p}=\frac{3(V_{r,un}-1)(5+2\Gamma[4+\Gamma])\cos\theta}{2(2+\Gamma)(2+3\Gamma)}\,.
\end{align}
Above,$V_{r,un}$ is the ratio of particle velocity in the uniform component to the maximum Poiseuille flow velocity; $V_{r,un}=V_{0,un}/V_{max}$. The surface deformation of the particle in the quadratic component is denoted by $f^{(1)}_{qad,p}$ and is obtained as
\begin{align}
    f^{(1)}_{qad,p}=& \frac{1}{64(2+\Gamma)(2+3\Gamma)}\left[\vphantom{\frac{1}{64(2+\Gamma)(2+3\Gamma)}}-64 P_0(2+\Gamma)+\cos\theta(96 V_{r,q} (5 + 2 \Gamma (4 + \right.\nonumber\\&\left.\Gamma))+ (260 + (136  - 21 \Gamma) \Gamma)\psi^2+ 35(2+\Gamma)(2+3\Gamma)\psi^2 (\cos2\theta\right.\nonumber\\&\left.-2\cos2\phi \sin^2\theta  )   )  \vphantom{\frac{1}{64(2+\Gamma)(2+3\Gamma)}}\right]\,.\label{appeq:f1 only in quadratic with Vr term}
\end{align}
Above, $V_{r,q}=V_{0,q}/V_{max}$. The surface deformation given in eqn (\ref{appeq:f1 only in quadratic with Vr term}) simplifies to that given in eqn (\ref{eq:f1 quadratic pp flow}) by substituting $V_{r,q}=1$ in eqn (\ref{appeq:f1 only in quadratic with Vr term}), which corresponds to the case of the particle translating at $V_{max}$ along the centreline in the quadratic component of the plane Poiseuille flow.

%\bibliography{Reference}% Produces the bibliography via BibTeX.
%\providecommand{\noopsort}[1]{}\providecommand{\singleletter}[1]{#1}%

%\makeatletter
%\renewcommand{\bibsection}{\@bibsection}
%\def\@bibsection{\section*{References}}  % removes horizontal rule
%\makeatother
%
\end{document}